\documentclass{iopjournal}

\usepackage{amsmath}
\usepackage{amssymb}
\usepackage{bm}

\usepackage{graphicx}
\usepackage{booktabs}
\usepackage{multirow}

\usepackage{siunitx}

\graphicspath{{figures/}}

\newcommand{\semg}{sEMG}

\begin{document}

\articletype{Paper}

\title{Material-Timescale Matching in Memristive Spiking Neural Networks for Continual Myoelectric Decoding}

\author{Ivan Kipelkin$^{1,2,3\ast}$}

\affil{$^1$ Laboratory of Stochastic Multistable Systems,
National Research Lobachevsky State University,
Nizhny Novgorod 603022, Russia\\
$^2$ Laboratory of Cognitive Modeling and Neurorobotics, Moscow Center for Advanced Studies, Moscow 123592, Russia\\
$^3$AI Centre, Skolkovo Institute of Science and Technology, Moscow, 121205, Russia.}

\affil{$^\ast$Author to whom any correspondence should be addressed.}

\email{kipelkinivan@gmail.com}



\begin{abstract}

\textit{Objective.}
Surface electromyographic decoders operate under persistent
non-stationarity caused by changes in electrode placement, muscle recruitment, fatigue, and recording conditions. We investigated whether the memory timescale of a memristive synapse should be selected relative to the timescale over which the input distribution changes, rather than treated as an isolated device property.

\textit{Approach.}
Six literature-constrained organic and inorganic memristive synapse models were implemented in the same feedforward spiking neural network and evaluated on the ten-session NinaPro DB6 dataset using chronological continual adaptation. Device-specific programming dynamics, conductance relaxation, and retention were preserved. Synaptic memory and controlled surface electromyographic drift were then varied independently over a nine-by-nine timescale grid and compared through
their relative memory-to-drift timescale. A hybrid synapse combining fast Ioannou and persistent Sahu device dynamics was evaluated under the same protocol.

\textit{Main results.}
Stationary decoding accuracy did not predict continual-learning behavior. Across the controlled sweep, continual-learning balanced accuracy formed a broad maximum when synaptic memory and input drift occurred on comparable timescales, reaching approximately 0.565 when memory was slightly faster than the drift process. A model based on the relative memory-to-drift timescale achieved a leave-one-participant-out root-mean-square error of 0.0154, compared with 0.0328 and 0.0327 for models based on memory and drift timescales alone. The hybrid synapse increased continual-learning balanced accuracy to 0.516, compared with 0.439 for Ioannou and 0.459 for Sahu, while reducing forgetting relative to the fast component.

\textit{Significance.}
These results identify memory-to-drift timescale matching as a useful
device-level design variable for continual myoelectric decoding. They also show that combining fast and persistent physical memory can shift the stability--plasticity trade-off, although at increased hardware and programming cost.

\end{abstract}


\section{Introduction}
\label{sec:introduction}

Reliable decoding of human motor intent remains a central challenge in neural engineering. Physiological signals are increasingly used to generate control commands for upper-limb prostheses, rehabilitation systems, wearable devices, and other human-machine interfaces. Surface electromyography (\semg{}) is particularly attractive in this context because it provides direct information about muscle activation with comparatively simple sensing hardware \cite{atzori2014ninapro,li2021semg,ni2024survey}. Its close relation to intended movement has made \semg{} a widely studied signal for intuitive prosthetic hand control and gesture-based interfaces.

Performance obtained in a single recording session often degrades when the same interface is used again. The statistical properties of \semg{} depend on electrode placement, skin-electrode conditions, contraction intensity, muscle recruitment, fatigue, and other physiological or acquisition-related factors. Some changes develop within a session; others become more pronounced after the electrodes have been removed and repositioned. Palermo \textit{et al.} reported an average 27.03\% decrease in classification accuracy when training and testing were performed on acquisitions recorded at different times rather than within the same acquisition \cite{palermo2017repeatability}. More recent studies on NinaPro Database 6 (DB6) and recordings from upper-limb amputees showed that cross-session calibration can recover part of this loss \cite{yao2024calibration,zhang2024crosssession}. Other approaches have addressed the same problem through adaptation. Li \textit{et al.} considered domain shifts across sessions and subjects in a myoelectric interface \cite{li2024across}, whereas Guo \textit{et al.} developed an inter-session domain adaptation method for \semg{} gesture recognition \cite{guo2024intersession}. Inter-subject variability has also been studied using unsupervised domain adaptation \cite{wang2024intersubject,zhang2025vae} and transfer learning aimed at reducing user-specific retraining \cite{fratti2024multiscale,zhang2024federated}.

This places a different demand on the learning system. New recording conditions must be incorporated while previously acquired information remains available when it is still relevant. Rapid adaptation increases plasticity, but unrestricted updates can lead to catastrophic forgetting. Strong preservation of existing parameters improves stability at the cost of responsiveness to new conditions. The resulting stability-plasticity problem is central to continual learning and becomes particularly relevant when physiological data change over time \cite{parisi2019continual,delange2022continual}. Bai \textit{et al.} studied class-incremental learning for high-density \semg{} under restricted memory resources \cite{bai2024memoryfriendly}. Li \textit{et al.} examined closed-loop co-adaptation in a dynamic myoelectric interface \cite{li2025closedloop}. Both settings require adaptation to be evaluated across a sequence of changing conditions rather than after a single recalibration step.

Neuromorphic hardware approaches the same problem differently. Memory need not exist only as a numerical parameter stored separately from the computing element. In a memristive device, conductance depends on previous electrical activity and can therefore act directly as the synaptic state. Programming history, relaxation, retention, and stochastic switching then become part of the learning substrate \cite{martemucci2025ferroelectric,chandrasekaran2026memristor}. This principle has already been applied to myoelectric systems. Tiw \textit{et al.} demonstrated an end-to-end memristive hardware platform in which memristive devices performed spike encoding and synaptic computation, including real-time control from \semg{} signals \cite{tiw2026hmi}. Manna \textit{et al.} recently demonstrated a neuromorphic spiking neural network (SNN) for EMG-based gesture classification, using resonate-and-fire neurons to enable low-power event-driven processing \cite{manna2026emg}. These studies establish the applicability of memristive neuromorphic processing to myoelectric interfaces, but they do not identify which physical memory dynamics are preferable when the input distribution continues to change after deployment.

Physical synaptic dynamics have also become relevant in continual-learning hardware more broadly. Zohora \textit{et al.} demonstrated that probabilistic metaplasticity implemented with low-precision memristive weights can reduce catastrophic forgetting in spiking networks without high-precision auxiliary memory \cite{zohora2024metaplasticity}. Chen \textit{et al.} demonstrated electrochemical ohmic memristors for continual learning and linked device-level conductance evolution directly to sequential adaptation \cite{chen2025electrochemical}. Ming \textit{et al.} used second-order memristors with slowly evolving internal states for continual reinforcement learning in dynamic environments \cite{ming2026secondorder}. In each case, temporal evolution of the physical device state participates in learning rather than appearing only as a non-ideal hardware effect.

However, a generic memristor model cannot capture this diversity. Organic and inorganic devices may implement synaptic conductance through different physical mechanisms. Organic iontronic and related memristive systems can involve ionic redistribution, charge trapping, redox processes, electrochemical modification, and coupled ionic-electronic transport, giving rise to a broad range of synaptic responses \cite{xia2024organic,gao2025organic,wang2026organic}. Metal-oxide memristors are commonly associated with oxygen-vacancy redistribution, interfacial redox processes, and the formation or rupture of conductive paths \cite{tian2025oxide,xia2025lowpower}. Their conductance states can persist for very different periods and differ substantially in terms of update nonlinearity, asymmetry, variability, and energy requirements.

Recent studies of volatile and adaptive memristive devices make these temporal differences especially relevant. Moon \textit{et al.} reviewed volatile memristors from material mechanisms to system-level neuromorphic implementation and emphasized that transient dynamics depend on the application rather than representing an intrinsically undesirable property \cite{moon2024leveraging,ju2024volatile}. Islam \textit{et al.} demonstrated a Li-imbued TiO$_x$ iontronic memristor whose volatile memory dynamics could be tuned through the excitation protocol, while the same platform could also operate in a non-volatile regime \cite{islam2024iontronic}. Chen \textit{et al.} further identified device dynamics, synaptic implementation, peripheral circuitry, and integration strategy as coupled constraints in scalable memristive spiking hardware \cite{chen2025scalable}. Therefore, a useful memory timescale cannot be assigned to at material system alone.

The organic-inorganic distinction remains useful for describing device technology, but it does not by itself explain network behaviour. Organic devices cannot generally be identified with short-term memory, just as inorganic devices cannot universally be treated as ideal non-volatile storage. Relaxation time, retention, conductance range, update characteristics, and stochasticity vary within both groups. For a decoder operating under temporal drift, these measured device properties may be more informative than the material label.

Different memory dynamics also raise the possibility that rapid adaptation and persistent storage can be carried by separate physical states. Dang \textit{et al.} combined an interfacial dynamic memristor exhibiting short-term fatigue with a hafnia-based non-volatile memristor, forming a hybrid synaptic element in which fast dynamics and persistent memory coexist within a SNN \cite{dang2026hybrid}. The architecture demonstrates that transient and consolidated synaptic states can be separated physically. Its value for a neural interface repeatedly exposed to changing input statistics remains unresolved.

In this paper, we consider two temporal processes: persistence of the memristive synaptic state and rate at which the myoelectric input distribution changes. A rapidly relaxing synapse may accommodate recent changes but lose information that remains useful across sessions. A highly persistent state may preserve previously learned structure while responding more slowly after recording condition changes. We test whether the relation between these processes determines the stability-plasticity behaviour of a continually adapting myoelectric decoder. Literature-constrained models derived from multiple organic and inorganic memristive devices are implemented in the same SNN under an identical learning protocol. The networks are evaluated on sequential multi-session \semg{} recordings together with controlled temporal drift. Stationary decoding, adaptation, retention, forgetting, backward transfer, and programming burden are compared across device models. Synaptic memory and input drift are then varied independently to determine whether their relative temporal scale describes continual-learning behaviour more consistently than either timescale alone. Finally, a hybrid fast-persistent implementation is examined to determine whether combining distinct physical memory dynamics changes the adaptation-retention trade-off.

\section{Methods}

\subsection{Study design and dataset}
\label{sec:study_design}

\begin{figure*}[ht!]
    \centering
    \includegraphics[width=0.96\textwidth]{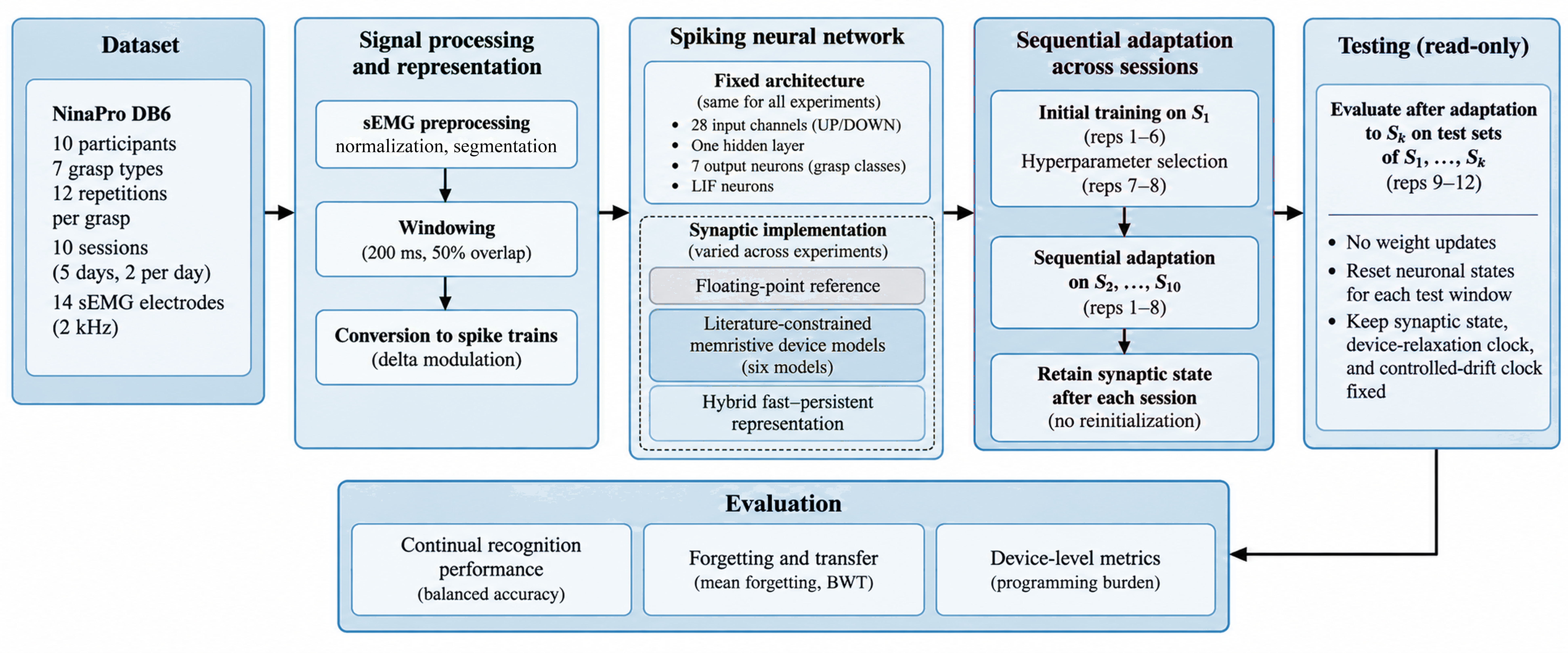}
    \caption{
Overview of the experimental framework for continual sEMG decoding.
NinaPro DB6 recordings from 10 participants performing seven grasp classes are segmented into 200~ms windows and converted into UP/DOWN spike trains by delta modulation. The resulting 28 input channels are supplied to a fixed feedforward SNN. Synapses are implemented using a floating-point reference, one of six literature-constrained memristive device models, or a hybrid fast--persistent representation. Initial training and hyperparameter selection use Session $S_1$, after which the network is adapted sequentially to $S_2,\ldots,S_{10}$ without synaptic reinitialization. Testing is read-only and leaves the synaptic state and internal device clocks unchanged. Evaluation includes balanced decoding accuracy, forgetting, backward transfer, and device-level programming burden.}
    \label{fig:study_design}
\end{figure*}

The paper examined how experimentally constrained synaptic dynamics influence continual adaptation to non-stationary myoelectric signals. The signal representation, network architecture, learning rule, data splits, and chronological session order were kept identical across experiments. Synaptic weights were represented either by floating-point variables, by one of six literature-constrained memristive device models, or by the hybrid fast--persistent implementation described in Section~\ref{sec:hybrid_hardware}. Figure~\ref{fig:study_design} summarizes the complete workflow from multi-session sEMG processing to sequential adaptation and read-only evaluation.

NinaPro DB6 was used as the primary dataset \cite{palermo2017repeatability}. The dataset contains recordings from 10 intact participants performing seven grasp types, with 12 repetitions of each grasp in every acquisition. Recordings were collected twice per day for five consecutive days, giving 10 acquisition sessions per participant. Muscle activity was sampled at 2~kHz from 14 Delsys Trigno wireless sEMG electrodes arranged in two circumferential groups: eight electrodes at the radio-humeral level and six positioned below them. The sessions were ordered by acquisition day and time and denoted $S_1,\ldots,S_{10}$. A separate decoder was trained for each participant. In $S_1$, repetitions 1--6 were used for initial training, repetitions 7--8 for validation, and repetitions 9--12 for testing. For $S_2,\ldots,S_{10}$, repetitions 1--8 formed the adaptation set, while repetitions 9--12 remained reserved for testing. Hyperparameters were selected only from $S_1$. After initial training, the synaptic state was retained while the decoder was adapted chronologically to $S_2,\ldots,S_{10}$. Before adaptation to session $S_k$ ($k \geq 2$), the current network state was evaluated on that session's reserved test set under read-only conditions. The same test set was evaluated again after adaptation. The resulting adaptation gain was

\begin{equation}
\Delta BA_{\mathrm{adapt},k}
=
BA^{\mathrm{post}}_k
-
BA^{\mathrm{pre}}_k,
\end{equation}

where $BA^{\mathrm{pre}}_k$ and $BA^{\mathrm{post}}_k$ denote balanced accuracy on the test set of session $S_k$ immediately before and after adaptation, respectively.

After adaptation to $S_k$, the same network state was also evaluated on the reserved test sets from $S_1,\ldots,S_k$. Neuronal states were reset for each test window, but the synaptic state, device-relaxation clock, and controlled-drift clock remained fixed. Testing therefore did not modify the state used for subsequent adaptation.

Natural inter-session variability in DB6 provided the application-level continual-learning condition. The controlled drift experiment described in Section~\ref{sec:timescale_matching} was performed separately.

\subsection{Signal preprocessing and spike encoding}
\label{sec:preprocessing_encoding}

The refined DB6 movement and repetition labels, \texttt{restimulus} and
\texttt{rerepetition}, were used to identify grasp intervals and
repetitions. For each participant and recording channel, the mean and
standard deviation were estimated from repetitions 1--6 of $S_1$ and
then retained for all later sessions. The normalized signal was

\begin{equation}
\tilde{x}_c(t)
=
\frac{x_c(t)-\mu_c^{(1)}}
{\sigma_c^{(1)}+\epsilon_{\mathrm{norm}}},
\label{eq:semg_normalization}
\end{equation}

where $\epsilon_{\mathrm{norm}}=10^{-8}$ is a numerical correction.
Normalization statistics were not recomputed after $S_1$.

Each contiguous labelled grasp interval was divided into 200~ms
windows with a 100~ms stride. At 2~kHz, each window contained 400
samples and consecutive windows were separated by 200 samples. Only
windows lying completely within a labelled grasp interval were kept.
Repetition-level splits were applied before segmentation, preventing
overlapping windows from crossing training, validation, adaptation,
or test boundaries.

Continuous sEMG was converted into events by delta modulation
\cite{ceolini2020handgesture}. For channel $c$, UP and DOWN events at
sample $k$ were defined as

\begin{equation}
\begin{aligned}
s_c^{+}[k] &=
\begin{cases}
1, & \tilde{x}_c[k]-r_c[k-1]\geq\theta,\\
0, & \text{otherwise},
\end{cases}
\\[4pt]
s_c^{-}[k] &=
\begin{cases}
1, & \tilde{x}_c[k]-r_c[k-1]\leq-\theta,\\
0, & \text{otherwise}.
\end{cases}
\end{aligned}
\label{eq:delta_events}
\end{equation}

The internal reference was updated as

\begin{equation}
r_c[k]
=
r_c[k-1]
+
\theta s_c^{+}[k]
-
\theta s_c^{-}[k],
\label{eq:delta_reference}
\end{equation}

with $r_c[0]=\tilde{x}_c[0]$ at the start of each window. The two event
conditions are mutually exclusive, so a channel could emit at most one
event within a single 0.5~ms sampling interval.

Candidate encoding thresholds were obtained from the pooled
distribution of absolute sample-to-sample increments across all
channels and training repetitions of $S_1$. Percentile levels of 70,
75, 80, 85, 90, and 95 were considered. For a global percentile $q$,
the corresponding threshold $\theta_p(q)$ was calculated separately
for each participant $p$. The percentile level was selected jointly
with hidden-layer size and learning rate as described in
Section~\ref{sec:snn_continual}. Once selected, the
participant-specific threshold remained fixed.

The 14 recording channels each produced one UP and one DOWN event
stream, resulting in 28 input spike channels.

\subsection{SNN and continual learning}
\label{sec:snn_continual}

The event streams were supplied to a feedforward SNN with one hidden layer, 28 input channels, and seven output neurons. Both
input-to-hidden and hidden-to-output connections were trainable. The network contained no recurrent connections or bias terms. Hidden and output neurons followed deterministic leaky integrate-and-fire dynamics. To keep logical synaptic weights dimensionless while retaining membrane potentials in millivolts, the synaptic drive was written as

\begin{equation}
\tau_{\mathrm{m}}
\frac{dU_i(t)}{dt}
=
-\left[U_i(t)-U_{\mathrm{rest}}\right]
+
J_i(t),
\label{eq:lif_voltage}
\end{equation}

\begin{equation}
\tau_{\mathrm{syn}}
\frac{dJ_i(t)}{dt}
=
-J_i(t)
+
V_{\mathrm{syn}}
\sum_j w_{ij}S_j(t),
\label{eq:synaptic_drive}
\end{equation}

where $w_{ij}\in[-1,1]$ denotes the dimensionless logical synaptic
weight. The fixed scale $V_{\mathrm{syn}}=10$~mV was equal to the
difference between the firing threshold and resting potential.

The neuronal parameters were
$\tau_{\mathrm{m}}=10$~ms,
$\tau_{\mathrm{syn}}=5$~ms,
$U_{\mathrm{rest}}=-60$~mV,
$\vartheta=-50$~mV, and a 5~ms refractory interval
\cite{zenke2018superspike}. After a spike, the membrane potential was
clamped at $U_{\mathrm{rest}}$ for the duration of the refractory
interval. No explicit axonal transmission delay was included.

Network dynamics were evaluated on the native 0.5~ms input grid.
First-order neuronal and filtering dynamics were integrated with
exponential Euler updates. Membrane potential, synaptic drive,
presynaptic traces, and eligibility traces were reset at the start of
each input window. Trainable synaptic states were carried across
windows and sessions.

Logical weights were initialized independently in each layer using Xavier uniform initialization \cite{glorot2010understanding},

\begin{equation}
w_{ij}(0)
\sim
\mathcal{U}
\left[
-\sqrt{\frac{6}{N_{\mathrm{in}}+N_{\mathrm{out}}}},
\sqrt{\frac{6}{N_{\mathrm{in}}+N_{\mathrm{out}}}}
\right],
\label{eq:weight_initialization}
\end{equation}

where $N_{\mathrm{in}}$ and $N_{\mathrm{out}}$ are the fan-in and
fan-out of the corresponding layer. Random seeds 1, 2, and 3 were used
for all reported simulations. For each participant and seed, every
synaptic implementation started from the same initial logical weights.

Synaptic adaptation followed a supervised three-factor rule based on a
surrogate derivative and a local eligibility trace
\cite{zenke2018superspike, bellec2020eprop}. Presynaptic activity was filtered with a
causal double-exponential kernel with rise and decay constants of
5~ms and 10~ms. The surrogate derivative was

\begin{equation}
\sigma'(U_i)
=
\frac{1}
{\left[
1+
\left|
\beta_{\mathrm{SG}}(U_i-\vartheta)
\right|
\right]^2},
\label{eq:superspike_surrogate}
\end{equation}

with $\beta_{\mathrm{SG}}=1~\mathrm{mV}^{-1}$. The local eligibility
trace was

\begin{equation}
e_{ij}(t)
=
\alpha
\ast
\left[
\sigma'(U_i(t))
\left(
\varepsilon\ast S_j
\right)(t)
\right],
\label{eq:eligibility_trace}
\end{equation}

where $\varepsilon$ denotes the presynaptic double-exponential kernel,
$\alpha$ is a second causal double-exponential filter, and $\ast$
denotes convolution. The eligibility filter used the same 5~ms rise
and 10~ms decay constants.

Output spikes were accumulated over the full 200~ms window. If $n_o$
is the number of spikes generated by output neuron $o$, the class
probabilities and loss were

\begin{equation}
p_o
=
\frac{\exp(n_o)}
{\sum_{r=1}^{N_{\mathrm{out}}}\exp(n_r)},
\qquad
\mathcal{L}
=
-\sum_{o=1}^{N_{\mathrm{out}}}
y_o\log p_o.
\label{eq:snn_classification_loss}
\end{equation}

The output learning signal was $\delta_o=p_o-y_o$. For hidden neuron
$i$, the corresponding signal was

\begin{equation}
\delta_i
=
\sum_{o=1}^{N_{\mathrm{out}}}
w_{oi}\delta_o.
\label{eq:hidden_learning_signal}
\end{equation}

One weight update was requested after each input window. With
$N_t=400$ simulation steps per window, the requested change was

\begin{equation}
\Delta w_{ij}^{*}
=
-\eta
\frac{1}{N_t}
\sum_{k=1}^{N_t}
\delta_i e_{ij}[k].
\label{eq:desired_synaptic_update}
\end{equation}

For output-layer connections, $\delta_i$ in
Equation~(\ref{eq:desired_synaptic_update}) denotes the corresponding
output error $\delta_o$. Averaging over the window prevents the update
magnitude from depending arbitrarily on the numerical integration time
unit.

For the floating-point reference, the requested update was applied
directly and the resulting weight was clipped to $[-1,1]$. Device
implementations converted the same requested update into programming
events as described in Section~\ref{sec:memristive-models}.

The encoding percentile, hidden-layer size, and learning rate were
selected jointly using the floating-point reference network.
Candidate hidden-layer sizes were 64, 128, and 256 neurons. Candidate
learning rates were $10^{-4}$, $5\times10^{-4}$, $10^{-3}$,
$5\times10^{-3}$, and $10^{-2}$. Each candidate configuration was
trained for exactly 20 epochs on repetitions 1--6 of $S_1$ and
evaluated on repetitions 7--8. Validation balanced accuracy was
averaged across the 10 participants and three matched initializations.
The configuration with the highest mean value was retained for all
subsequent experiments.

The selected values were $N_h=256$, $\eta=5\times10^{-4}$, and
$q=85$. They remained fixed for the stationary, continual-learning,
controlled-timescale, and hybrid experiments. No early stopping or
device-specific hyperparameter optimization was used. The complete
hyperparameter search and participant-level validation results for the
selected configuration are provided in Supplementary Table~S1.

Training windows were randomly permuted at the beginning of every
$S_1$ epoch using the corresponding simulation seed. For each later
session, the adaptation set was presented once in a fixed random
permutation generated from the same seed. The window order was shared
across all synaptic implementations. Each window was assigned to the output neuron with the largest spike count. Windows producing no output spikes were counted as incorrect, while non-zero ties were resolved by the earliest output spike. No replay, rehearsal, elastic-weight regularization, or software metaplasticity was used during $S_2,\ldots,S_{10}$.

\subsection{Literature-constrained memristive synapse models}
\label{sec:memristive-models}

Six experimentally characterized two-terminal synaptic devices were used to construct the device-constrained synaptic models. The PEDOT:PSS/pentacene device reported by Luo \textit{et al.} \cite{luo2022organic}, the Ta/EV(ClO$_4$)$_2$/BTPA-F/Pt device of Zeng \textit{et al.} \cite{zeng2023organic}, and the PEDOT:PSS synaptic wire of Watanabe \textit{et al.} \cite{watanabe2024pedot} were assigned to the organic group. The inorganic group comprised the Ag/TiO$_2$--WO$_{3-x}$/ITO/PET device of Pan \textit{et al.} \cite{pan2024tio2wo}, the Au/Li$_x$CoO$_2$/SiO$_x$/TiO$_2$/p$^{++}$-Si device of Ioannou \textit{et al.} \cite{ioannou2020libased}, and the Ti/HfO$_2$/HfO$_{2-x}$/Pt device of Sahu \textit{et al.} \cite{sahu2023hfox}.

The measurements retained for device parameterization are listed in Table~\ref{tab:device-parameterization}. Potentiation refers to an increase in the measured conductance or conductance-proportional state variable, while depression refers to a decrease, regardless of programming-voltage polarity.

\begin{table*}[t]
\centering
\caption{Experimental measurements used to construct the six literature-constrained synaptic device models.}
\label{tab:device-parameterization}
\small
\begin{tabular}{p{3.2cm} p{1.5cm} p{2.0cm} p{4.2cm} p{3.0cm}}
\toprule
Device & Class & Programming source & Programming protocol & Memory source \\
\midrule

PEDOT:PSS/pentacene
\cite{luo2022organic}
&
Organic
&
Figure~3A
&
50 pulses at $+2.0$~V followed by 50 pulses at $-1.5$~V;
100~ms width and 100~ms interval
&
Post-stimulation relaxation in Figure~2
\\

Ta/EV(ClO$_4$)$_2$/BTPA-F/Pt
\cite{zeng2023organic}
&
Organic
&
Figure~4a
&
50 pulses at $+1.0$~V followed by 50 pulses at $-1.0$~V;
10~ms width and 2~s period
&
Retention in Figure~2c
\\

PEDOT:PSS wire
\cite{watanabe2024pedot}
&
Organic
&
Figure~3c
&
50-pulse sequences at $\pm1.0$~V;
10~ms width and 800~ms interval
&
Post-stimulation relaxation in Figure~4a
\\

Ag/TiO$_2$--WO$_{3-x}$/ITO/PET
\cite{pan2024tio2wo}
&
Inorganic
&
Figure~2h
&
Potentiation and depression at $\pm1.0$~V;
30~ms width and 30~ms interval
&
Post-stimulation decay in Figure~2g
\\

Au/Li$_x$CoO$_2$/SiO$_x$/\\TiO$_2$/p$^{++}$-Si
\cite{ioannou2020libased}
&
Inorganic
&
Figure~4
&
Potentiation at $-4.2$~V and depression at $+4.2$~V;
60~ms width and 120~ms spacing
&
Relaxation in Figure~4 and Supplementary Information
\\

Ti/HfO$_2$/HfO$_{2-x}$/Pt
\cite{sahu2023hfox}
&
Inorganic
&
Figure~8
&
Identical square pulses at $\pm2.0$~V;
10~$\mu$s width and 1~$\mu$s interval
&
Retention in Figures~2e and 3b
\\

\bottomrule
\end{tabular}
\end{table*}

Machine-readable author source data were used when available. Otherwise, the selected curves were reconstructed from published figures by calibrated raster digitization. For device $d$, the measured programming observable is denoted by $x_d$. It corresponds to conductance for Luo, Zeng, Watanabe, Ioannou, and Sahu. In the Pan data, Figure~2h reports current without a panel-specific read voltage, so this quantity was treated only as a conductance-proportional programming variable. Each device was normalized independently as

\begin{equation}
g_d
=
\frac{x_d-x_{\min,d}}
{x_{\max,d}-x_{\min,d}},
\qquad
0\leq g_d\leq1,
\label{eq:normalized-conductance}
\end{equation}

where $x_{\min,d}$ and $x_{\max,d}$ were taken from the combined experimentally supported potentiation and depression range. Absolute conductance and current scales were not compared across devices. Source-specific data provenance, reconstruction procedures, and native memory representations are summarized in Supplementary Table~S2.

A common parametric programming law was first tested as a diagnostic description of the state dependence of potentiation and depression,

\begin{equation}
F_{+,d}(g)
=
A_{+,d}
(1-g)^{p_{+,d}}
(g+\varepsilon_F)^{q_{+,d}},
\label{eq:memristive-potentiation}
\end{equation}

\begin{equation}
F_{-,d}(g)
=
A_{-,d}
g^{p_{-,d}}
(1-g+\varepsilon_F)^{q_{-,d}},
\label{eq:memristive-depression}
\end{equation}

with $\varepsilon_F=10^{-6}$,
$10^{-6}\leq A_{\pm,d}\leq1$, and
$0\leq p_{\pm,d},q_{\pm,d}\leq10$.
The modeled state was then propagated recursively,

\begin{equation}
\widetilde g_{n+1}
=
\begin{cases}
\mathcal{C}\!\left[
\widetilde g_n+F_{+,d}(\widetilde g_n)
\right],
& \text{potentiation},\\[4pt]
\mathcal{C}\!\left[
\widetilde g_n-F_{-,d}(\widetilde g_n)
\right],
& \text{depression},
\end{cases}
\label{eq:parametric-programming-update}
\end{equation}

where $\mathcal{C}(x)=\min[1,\max(0,x)]$. Complete trajectories were fitted by least-squares minimization. Fit quality was assessed from the trajectory root-mean-square error (RMSE), $R^2$, residual structure, and convergence toward the imposed parameter bounds. The common parametric form did not reproduce every experimentally characterized branch with comparable fidelity. The same empirical representation was therefore used for all six devices.

Programming was represented by a device-specific shape-preserving piecewise cubic Hermite interpolant (PCHIP) \cite{fritsch1980monotone},

\begin{equation}
g_{\pm,d}(\xi)
=
\mathcal{P}_{\pm,d}(\xi),
\label{eq:pchip-programming}
\end{equation}

where $\xi$ is the experimentally supported programming coordinate. For discrete protocols, $\xi$ corresponds to pulse index. The continuously digitized pulse coordinate was retained for Pan. When repeated measurements were available, values at matching programming coordinates were aggregated before interpolation. Luo states were averaged across repeated half-cycles, Watanabe states across the available electrolyte and cycle realizations, and the Ioannou trajectory was represented by the five-cycle mean. Single selected branches were retained for Zeng, Pan, and Sahu.

No smoothing or isotonic correction was applied before interpolation. Local reversals present in the source data were retained. Backward extrapolation was not permitted, and the final experimentally supported state was held beyond the terminal knot. Each trajectory was checked for consistency at the source knots, normalized-state bounds, and PCHIP overshoot.

For device-level comparison, the normalized branch range was defined as

\begin{equation}
R_{\pm,d}
=
\max_{\xi} g_{\pm,d}(\xi)
-
\min_{\xi} g_{\pm,d}(\xi).
\label{eq:branch-range}
\end{equation}

For experimentally discrete pulse coordinates, the local programming increment was

\begin{equation}
\delta g_{\pm,d}(n)
=
g_{\pm,d}(n+1)-g_{\pm,d}(n).
\label{eq:programming-increment}
\end{equation}

Programming granularity was characterized from the distribution of $|\delta g_{\pm,d}|$, including its median and percentile range. Because the Pan programming coordinate was reconstructed as a continuous variable, its PCHIP trajectory was also evaluated on the supported integer-pulse grid for the cross-device single-pulse comparison.

Programming asymmetry was quantified as

\begin{equation}
A_{\mathrm{sym},d}
=
\frac{
\operatorname{median}|\delta g_{+,d}|
-
\operatorname{median}|\delta g_{-,d}|
}{
\operatorname{median}|\delta g_{+,d}|
+
\operatorname{median}|\delta g_{-,d}|
},
\label{eq:programming-asymmetry}
\end{equation}

provided that the denominator was non-zero. Positive values correspond to larger characteristic potentiating increments, while negative values correspond to larger depressing increments. A local direction violation was recorded whenever the sign of a neighboring increment opposed the nominal global direction of its programming branch.

Signed logical weights were implemented with differential conductance pairs,

\begin{equation}
w_{ij}
=
g_{ij}^{+}
-
g_{ij}^{-},
\qquad
-1\leq w_{ij}\leq1,
\label{eq:differential-synapse}
\end{equation}

initialized as

\begin{equation}
g_{ij}^{+}(0)
=
\frac{1+w_{ij}(0)}{2},
\qquad
g_{ij}^{-}(0)
=
\frac{1-w_{ij}(0)}{2}.
\label{eq:differential-initialization}
\end{equation}

A positive requested logical update potentiated $g_{ij}^{+}$ and depressed $g_{ij}^{-}$; a negative update applied the complementary operation. Each physical element retained its current normalized state and its position on the active programming branch. If the programming direction changed, the current state was mapped to the nearest supported state on the target branch. This changed the branch coordinate without changing conductance.

For a requested update $\Delta w_{ij}^{*}$, candidate integer pulse counts were evaluated directly from the empirical programming trajectories. After $m$ effective differential programming steps, the resulting weight change was

\begin{equation}
\Delta w_{ij}^{(d)}(m)
=
w_{ij}^{(d)}(m)-w_{ij}^{(d)}(0).
\label{eq:empirical-weight-change}
\end{equation}

The pulse count was selected according to

\begin{equation}
m_{ij}^{*}
=
\underset{m}{\arg\min}
\left|
\Delta w_{ij}^{*}
-
\Delta w_{ij}^{(d)}(m)
\right|,
\label{eq:pulse-mapping}
\end{equation}

with the candidate set restricted to the experimentally supported programming domain. When two pulse counts gave the same approximation error, the smaller value was selected. One effective differential step applies one physical pulse to each member of the differential pair, so an update of $m$ steps contributes $2m$ physical programming pulses.

The pulse widths and intervals listed in Table~\ref{tab:device-parameterization} are already contained in the measured programming trajectories and were not added separately to the global network clock.

Native memory dynamics were represented according to what could be identified from each source experiment. If a reference level and a finite relaxation profile could be resolved, free conductance evolution was described by

\begin{equation}
g_d(t)
=
g_{b,d}
+
\left[g_d(t_0)-g_{b,d}\right]
\left[
\rho_d
+
(1-\rho_d)
\exp\left(
-
\left[
\frac{t-t_0}{\tau_d}
\right]^{\beta_d}
\right)
\right],
\label{eq:parametric-relaxation}
\end{equation}

with
$10^{-3}\leq\tau_d\leq10^{6}$~s,
$0.2\leq\beta_d\leq2.5$, and
$0\leq\rho_d\leq1$.
The reference level $g_{b,d}$ was used only when supported by the source-specific normalization. It was not treated as an independently measured equilibrium conductance.

Some source data contained several state-, history-, or condition-dependent relaxation trajectories without an identifiable common asymptote. In those cases, the relaxation response was retained directly in empirical form,

\begin{equation}
y_{d,r}(t)
=
\mathcal{R}_{d,r}(t),
\label{eq:empirical-relaxation}
\end{equation}

where $\mathcal{R}_{d,r}$ is a shape-preserving PCHIP representation of measured trajectory $r$. Backward time extrapolation was prohibited, and the final measured value was held after the end of the supported interval.

For Watanabe, relaxation trajectories were kept separate by electrolyte and initial state, with interpolation restricted to measured states within the same electrolyte. The five Pan trajectories corresponding to 20, 50, 100, 150, and 200 preceding stimulation pulses were retained in the native Figure~2g representation because no experimentally established absolute mapping to the Figure~2h programming scale was available. Ioannou relaxation trajectories were retained by source family and initial state, again with interpolation restricted to measured states within the same family.

No finite $\tau_d$, $\beta_d$, or $\rho_d$ was assigned when the reported retention interval did not contain a resolved monotonic decay. Such devices were treated as persistent only over the experimentally supported retention window.

For empirical transient trajectories, the relative observed change was

\begin{equation}
\Delta y_{\mathrm{rel}}
=
\frac{
y(t_{\mathrm{end}})-y(t_0)
}{
|y(t_0)|
}.
\label{eq:relative-observed-change}
\end{equation}

The fraction of the observed start-to-end change was defined as

\begin{equation}
f_{\mathrm{obs}}(t)
=
\frac{
y(t)-y(t_0)
}{
y(t_{\mathrm{end}})-y(t_0)
}.
\label{eq:observed-change-fraction}
\end{equation}

The observed-change descriptors $t_{50}$ and $t_{1/e}$ were the earliest times at which $f_{\mathrm{obs}}(t)$ reached $0.5$ and $1-e^{-1}$, respectively. They characterize the measured transition over the available observation window and were not interpreted as fitted physical time constants. If the corresponding crossing was absent or not uniquely defined, the descriptor was left undefined.

\subsection{Temporal drift and timescale-matching protocol}
\label{sec:timescale_matching}

Controlled input drift was introduced separately from the native inter-session variability of DB6. Electrode displacement provided the physical motivation, since spatial shifts of surface electrodes alter the multichannel sEMG distribution \cite{young2012electrodeshift}. 

A one-position cyclic displacement within each of the two electrode groups was represented by
$\mathbf{P}_{\delta}
=
\operatorname{diag}
(\mathbf{P}_{8,\delta},\mathbf{P}_{6,\delta})$,
with $\delta\in\{-1,+1\}$. For adaptation window $m$, the transformed input was

\begin{equation}
\tilde{\mathbf{x}}^{\,\mathrm{drift}}_m[k]
=
(1-\lambda_m)\tilde{\mathbf{x}}_m[k]
+
\lambda_m
\mathbf{P}_{\delta}\tilde{\mathbf{x}}_m[k],
\label{eq:controlled_channel_drift}
\end{equation}

where

\begin{equation}
\lambda_m
=
1-
\exp\left(
-\frac{t_m}{\tau_{\mathrm{drift}}}
\right).
\label{eq:drift_transition}
\end{equation}

The controlled drift started with the first adaptation window of $S_2$, for which $t_m=0$. Each subsequent adaptation-window presentation advanced $t_m$ by 200~ms. The clock continued across session boundaries but remained fixed during evaluation.

For each test session $S_j$, the value of $\lambda$ reached immediately after adaptation to that session was stored. The corresponding reserved test set was transformed once using this value and reused unchanged in all later evaluations of $S_j$. Performance changes on an earlier session therefore arose from changes in the network state, not from continued modification of its test distribution.

The drift sweep used
$\tau_{\mathrm{drift}}=0.2$, 0.8, 3.2, 12.8, 51.2, 204.8, 819.2,
3276.8, and 13107.2~s. Both displacement directions were evaluated.

For the controlled-memory experiment, all timescale conditions belonging to the same participant, device model, and random seed started from the same trained $S_1$ endpoint. This state was cloned before adaptation to $S_2$, so varying $\tau_{\mathrm{mem}}$ or $\tau_{\mathrm{drift}}$ did not alter the initial decoder.

Device-specific programming functions were retained, whereas native free relaxation was replaced by controlled decay of the differential weight. For a differential pair,

\begin{equation}
\bar{g}_{ij}(t_0)
=
\frac{
g_{ij}^{+}(t_0)+g_{ij}^{-}(t_0)
}{2},
\label{eq:pair_common_mode}
\end{equation}

and free evolution after the most recent programming event followed

\begin{equation}
g_{ij}^{\pm}(t)
=
\bar{g}_{ij}(t_0)
+
\left[
g_{ij}^{\pm}(t_0)-\bar{g}_{ij}(t_0)
\right]
\exp\left(
-\frac{t-t_0}{\tau_{\mathrm{mem}}}
\right).
\label{eq:controlled_memory_decay}
\end{equation}

This preserves the common-mode conductance while allowing the differential logical weight to decay with the controlled timescale $\tau_{\mathrm{mem}}$. Conductance relaxation was evaluated at every 0.5~ms adaptation step. Evaluation used a frozen synaptic snapshot, as in the native-device simulations, and did not advance the memory clock.

The memory sweep used
$\tau_{\mathrm{mem}}=0.2$, 0.8, 3.2, 12.8, 51.2, 204.8, 819.2,
3276.8, and 13107.2~s.

The relative timescale was defined as

\begin{equation}
\rho_{\tau}
=
\frac{\tau_{\mathrm{mem}}}
{\tau_{\mathrm{drift}}}.
\label{eq:timescale_ratio}
\end{equation}

\begin{figure*}[ht!]
    \centering
    \includegraphics[width=0.98\textwidth]{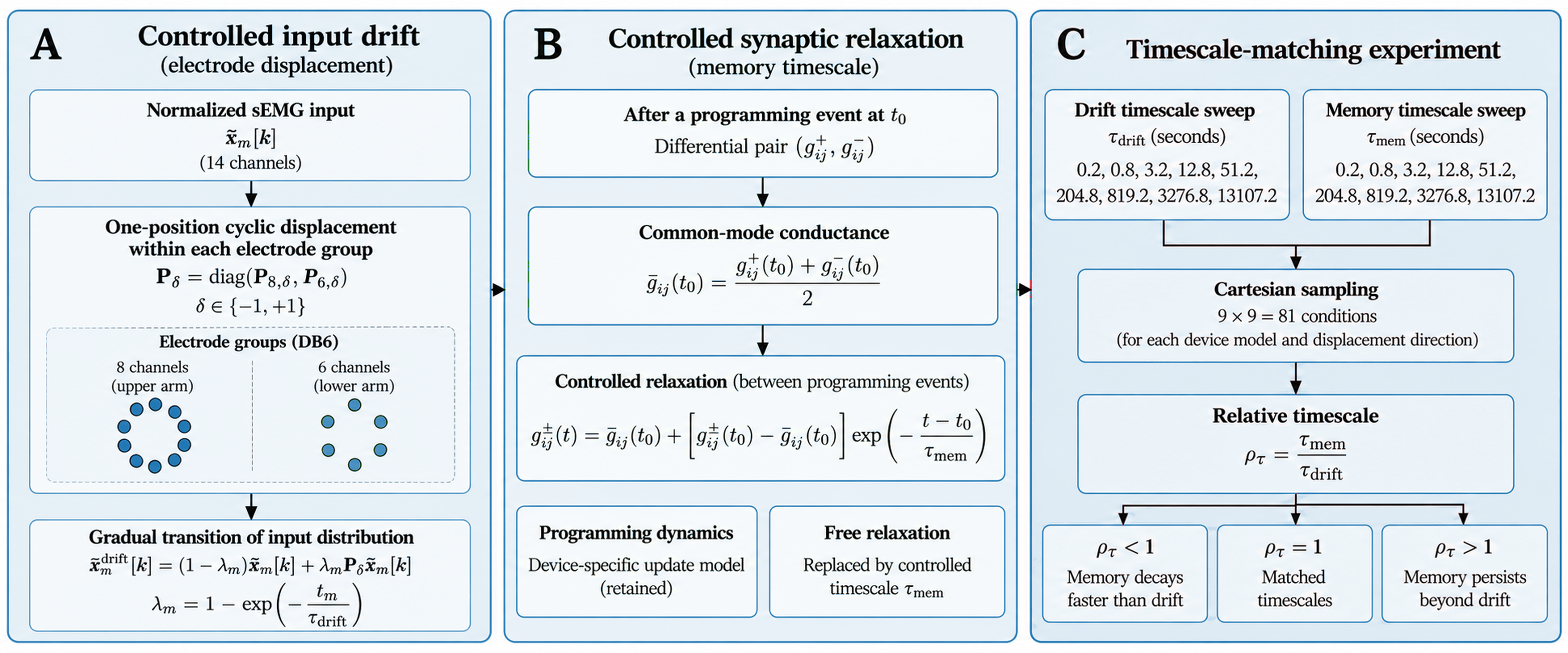}
    \caption{Controlled drift and synaptic-memory timescale protocol.
(A) Input-distribution drift is generated by a one-position cyclic displacement
within the two DB6 electrode groups and introduced progressively according to
the characteristic drift timescale $\tau_{\mathrm{drift}}$.
(B) During the controlled-memory experiment, device-specific programming
dynamics are retained, whereas native free relaxation is replaced by
exponential decay of the differential conductance around its common-mode
component with characteristic timescale $\tau_{\mathrm{mem}}$.
(C) Both timescales are sampled independently over the same nine-point
logarithmic grid, yielding $9\times9=81$ combinations for each device-update
model and displacement direction. Their relative scale is expressed as
$\rho_\tau=\tau_{\mathrm{mem}}/\tau_{\mathrm{drift}}$, with
$\rho_\tau<1$, $\rho_\tau=1$, and $\rho_\tau>1$ corresponding to memory that
decays faster than, matches, or persists longer than the imposed input drift,
respectively.}
    \label{fig:timescale_protocol}
\end{figure*}

Values $\rho_{\tau}<1$ indicate memory that decays faster than the imposed distributional drift, while $\rho_{\tau}>1$ indicates memory that persists beyond the characteristic drift time. The protocol is summarized in Figure~\ref{fig:timescale_protocol}. 

The Cartesian product of the nine memory times and nine drift times produced 81 controlled conditions for each literature-derived device-update model and displacement direction.

\subsection{Hybrid synapse and hardware-aware evaluation}
\label{sec:hybrid_hardware}

The primary hybrid combined the fast-relaxing Li-based device of Ioannou \textit{et al.} \cite{ioannou2020libased} with the persistent HfO$_x$ device of Sahu \textit{et al.} \cite{sahu2023hfox}. Each logical connection contained one differential pair for the fast branch and another for the persistent branch,

\[
d_{ij}^{\mathrm{f}}
=
g_{ij,\mathrm{f}}^{+}
-
g_{ij,\mathrm{f}}^{-},
\qquad
d_{ij}^{\mathrm{s}}
=
g_{ij,\mathrm{s}}^{+}
-
g_{ij,\mathrm{s}}^{-}.
\]

The effective logical weight was

\begin{equation}
w_{ij}
=
\frac{1}{2}d_{ij}^{\mathrm{f}}
+
\frac{1}{2}d_{ij}^{\mathrm{s}},
\label{eq:hybrid_weight}
\end{equation}

which preserves the range $[-1,1]$. Both differential pairs were initialized from the same logical $w_{ij}(0)$ using Equation~(\ref{eq:differential-initialization}).

No additional consolidation algorithm was introduced. For a requested update,

\begin{equation}
\Delta d_{ij}^{\mathrm{f},*}
=
\Delta d_{ij}^{\mathrm{s},*}
=
\Delta w_{ij}^{*}.
\label{eq:hybrid_requested_update}
\end{equation}

The requested change for each branch was converted independently using the pulse-mapping rule of the corresponding device. The realized change in the hybrid logical weight was then

\begin{equation}
\Delta w_{ij}^{\mathrm{actual}}
=
\frac{1}{2}
\Delta d_{ij}^{\mathrm{f,actual}}
+
\frac{1}{2}
\Delta d_{ij}^{\mathrm{s,actual}}.
\label{eq:hybrid_actual_update}
\end{equation}

Between programming events, each branch followed its own literature-constrained memory dynamics. No software transfer, consolidation threshold, branch-specific learning rate, or adaptive routing parameter was introduced.

Ioannou--Sahu was used as the primary hybrid pairing. Ioannou--Zeng, Pan--Sahu, and Pan--Zeng were evaluated under the same equations as sensitivity analyses \cite{zeng2023organic,pan2024tio2wo}. 

For hidden-layer size $N_h$, the network contains
$N_{\mathrm{syn}}=28N_h+7N_h=35N_h$ trainable logical synapses. A single-device implementation uses two physical conductance elements per logical synapse, while the hybrid uses four.

If $n_r$ denotes the cumulative number of programming pulses applied to physical device $r$, programming burden was summarized as

\begin{equation}
N_{\mathrm{pulse}}
=
\sum_{r=1}^{N_{\mathrm{dev}}}n_r,
\qquad
N_{\mathrm{pulse}}^{\max}
=
\max_r n_r.
\label{eq:programming_burden}
\end{equation}

Pulses applied to different members of a differential pair were counted separately. In hybrid simulations, pulse counts for the fast and persistent branches were also retained separately.
$N_{\mathrm{pulse}}^{\max}$ was treated as a relative device-level endurance burden rather than a failure prediction.

Programming energy was not compared across devices because the six source studies did not report a uniform set of programming-current or pulse-energy measurements.

\subsection{Evaluation metrics and statistical analysis}
\label{sec:evaluation_statistics}

Classification performance was measured using balanced accuracy,

\begin{equation}
\mathrm{BA}
=
\frac{1}{C}
\sum_{c=1}^{C}
\frac{TP_c}{TP_c+FN_c},
\qquad C=7.
\label{eq:balanced_accuracy}
\end{equation}

Let $BA_{k,\ell}$ denote balanced accuracy on the fixed test set of $S_\ell$ after adaptation through $S_k$, with $\ell \le k$.

For each adaptation stage $k$, mean retention across all previously encountered sessions was defined as
\begin{equation}
R_k=\frac{1}{k}\sum_{\ell=1}^{k} BA_{k,\ell}.
\end{equation}

Continual-learning balanced accuracy was defined as

\begin{equation}
\mathrm{BA}_{\mathrm{CL}}
=
\frac{1}{K}
\sum_{k=1}^{K}
\left(
\frac{1}{k}
\sum_{\ell=1}^{k}
\mathrm{BA}_{k,\ell}
\right),
\qquad K=10.
\label{eq:continual_accuracy}
\end{equation}

Forgetting was quantified by

\begin{equation}
F_\ell
=
\max_{k\in\{\ell,\ldots,K\}}
\mathrm{BA}_{k,\ell}
-
\mathrm{BA}_{K,\ell},
\qquad
\overline{F}
=
\frac{1}{K-1}
\sum_{\ell=1}^{K-1}F_\ell.
\label{eq:forgetting}
\end{equation}

Backward transfer was defined as

\begin{equation}
\mathrm{BWT}
=
\frac{1}{K-1}
\sum_{\ell=1}^{K-1}
\left(
\mathrm{BA}_{K,\ell}
-
\mathrm{BA}_{\ell,\ell}
\right).
\label{eq:backward_transfer}
\end{equation}

The complete $\mathrm{BA}_{k,\ell}$ matrix was retained for session-resolved analyses.

For each participant and condition, the three matched random initializations were averaged before inferential statistics were calculated. Participants were treated as the independent statistical units, rather than windows, device elements, or random seeds. Group-level descriptive statistics were reported as participant means, with 95\% bootstrap confidence intervals where applicable. 

The controlled-timescale experiment used $\mathrm{BA}_{\mathrm{CL}}$ as its primary response. Four predictive models were specified in advance:

\[
\begin{aligned}
M_{\mathrm{mem}} &:\
\mathrm{BA}_{\mathrm{CL}}
\sim
f(\log_{10}\tau_{\mathrm{mem}})
+
\mathrm{device}
+
\mathrm{direction},
\\
M_{\mathrm{drift}} &:\
\mathrm{BA}_{\mathrm{CL}}
\sim
f(\log_{10}\tau_{\mathrm{drift}})
+
\mathrm{device}
+
\mathrm{direction},
\\
M_{\rho} &:\
\mathrm{BA}_{\mathrm{CL}}
\sim
f(\log_{10}\rho_\tau)
+
\mathrm{device}
+
\mathrm{direction},
\\
M_{\mathrm{abs}} &:\
\mathrm{BA}_{\mathrm{CL}}
\sim
f(\log_{10}\tau_{\mathrm{mem}})
+
f(\log_{10}\tau_{\mathrm{drift}})
+
\mathrm{device}
+
\mathrm{direction}.
\end{aligned}
\]

Each $f(\cdot)$ was represented by a natural cubic spline with four degrees of freedom, with the same spline complexity used for every one-dimensional temporal predictor. Predictive performance was estimated by leave-one-participant-out cross-validation. In each fold, all conditions belonging to one participant were withheld during fitting. Cross-validated $R^2$ and root-mean-square error were calculated from the pooled held-out predictions. No assumption of timescale collapse was imposed during model fitting.


\section{Results}
\label{sec:results}

\subsection{Literature-constrained models reproduce distinct synaptic dynamics}
\label{sec:results-synapses}

Normalization to the common device-local interval $g\in[0,1]$ did not remove the pronounced differences in programming and memory dynamics among the six literature-constrained models. The diagnostic recursive model defined in Equation~(\ref{eq:memristive-potentiation}) and~(\ref{eq:memristive-depression}) described several programming trajectories reasonably well, but its fidelity was not comparable across all devices. The largest discrepancy occurred for Ioannou. After averaging the five experimental programming cycles, the potentiation trajectory reached $R^2=0.883$ with $\mathrm{RMSE}_g=0.0302$, whereas the depression trajectory gave only $R^2=0.132$ and $\mathrm{RMSE}_g=0.0440$. The depression fit also converged toward the imposed parameter boundaries. Following the model-selection procedure in Section~\ref{sec:memristive-models}, the empirical PCHIP representation was used as the primary programming model for all six devices.

\begin{figure*}[ht!]
    \centering
    \includegraphics[width=\textwidth]{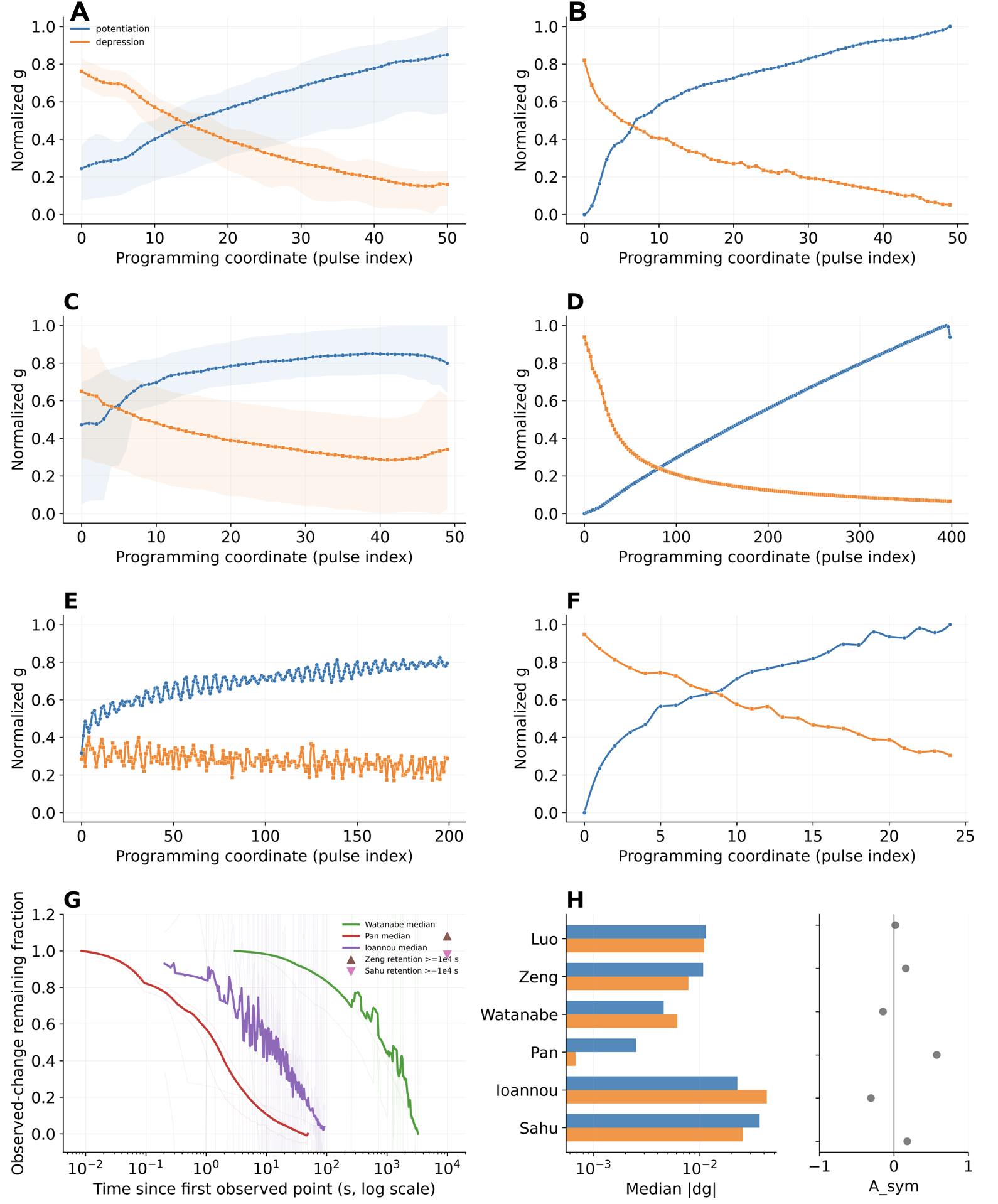}
    \caption{
    Literature-constrained programming and native memory characteristics of the six synaptic devices.
    (A) Luo \textit{et al.}: normalized potentiation and depression trajectories of the PEDOT:PSS/pentacene device.
    (B) Zeng \textit{et al.}: normalized potentiation and depression trajectories of the Ta/EV(ClO$_4$)$_2$/BTPA-F/Pt device.
    (C) Watanabe \textit{et al.}: programming trajectories of the PEDOT:PSS wire obtained by aggregation across the experimentally characterized electrolyte and cycle realizations.
    (D) Pan \textit{et al.}: programming trajectories of the Ag/TiO$_2$--WO$_{3-x}$/ITO/PET device; the conductance-proportional current reported in the source experiment is expressed in the device-local normalized state $g$.
    (E) Ioannou \textit{et al.}: five-cycle mean programming trajectories reconstructed from the machine-readable source data.
    (F) Sahu \textit{et al.}: fixed-amplitude potentiation and depression trajectories of the Ti/HfO$_2$/HfO$_{2-x}$/Pt device.
    In panels (A--F), points denote experimentally extracted or aggregated programming states and curves denote the corresponding shape-preserving PCHIP representations. No smoothing or isotonic correction was applied.
    (G) Native memory characteristics. Transient trajectories are expressed relative to their experimentally observed start-to-end change, while Zeng and Sahu are represented by their experimentally supported retention lower bound $T_{\mathrm{ret}}\geq10^{4}$~s.
    (H) Cross-device comparison of the median normalized single-step programming magnitude, the bounded potentiation--depression asymmetry index $A_{\mathrm{sym}}$, and the corresponding native memory descriptor. Absolute conductance and current scales are device specific and are not compared directly across devices.
    }
    \label{fig:device-characterization}
\end{figure*}

The final database contained 12 potentiation and depression trajectories, 1118 experimental or aggregated knots, and 1106 cubic PCHIP segments. Every trajectory reproduced the retained source points within numerical precision, remained inside the normalized state bounds, and passed the no-overshoot check. These tests verify interpolation consistency rather than providing an independent measure of goodness of fit. Figure~\ref{fig:device-characterization} summarizes the resulting device responses.

Luo exhibited closely matched potentiation and depression characteristics [Figure~\ref{fig:device-characterization}(A)]. The normalized branch ranges were $0.605$ and $0.611$, while the median single-step magnitudes were $0.01137$ for potentiation and $0.01096$ for depression. The resulting asymmetry index was $A_{\mathrm{sym}}=0.018$. No local direction violations occurred along the potentiation trajectory, and only $2.0\%$ of depression increments locally opposed the expected branch direction. 

Zeng also showed a regular potentiation trajectory, although the two programming directions differed more strongly [Figure~\ref{fig:device-characterization}(B)]. The median normalized step magnitudes were $0.01074$ for potentiation and $0.00782$ for depression, giving $A_{\mathrm{sym}}=0.158$. Potentiation remained monotonic over the reconstructed branch. By contrast, $8.2\%$ of depression increments showed local reversals. The potentiation branch covered the full normalized programming interval, whereas the depression branch terminated above the lower normalized boundary, preserving the asymmetry observed experimentally.

The Watanabe device operated in a finer update regime [Figure~\ref{fig:device-characterization}(C)]. Median potentiating and depressing step magnitudes were $0.00456$ and $0.00613$, respectively, corresponding to $A_{\mathrm{sym}}=-0.147$. The negative value reflects the larger characteristic depression step. Local reversals were more frequent than for Luo or Zeng, affecting $22.4\%$ of potentiation increments and $16.3\%$ of depression increments. These programming trajectories were obtained by aggregating the experimentally available states across six electrolyte conditions and four programming cycles, while electrolyte identity remained separate in the native relaxation analysis.

Pan provided the finest programming resolution of the six models [Figure~\ref{fig:device-characterization}(D)]. Its pulse coordinate was reconstructed as a continuous variable from the published programming panel, so the empirical PCHIP trajectory was also evaluated on the supported integer-pulse grid before single-pulse descriptors were calculated. Median normalized increments were $2.50\times10^{-3}$ for potentiation and $6.78\times10^{-4}$ for depression. This produced $A_{\mathrm{sym}}=0.574$, the largest positive asymmetry in the device set. About $1.0\%$ of potentiation increments locally opposed the nominal programming direction, while no direction violations were found for depression.

The Ioannou trajectories differed markedly from the smoother responses of the preceding devices [Figure~\ref{fig:device-characterization}(E)]. Median step magnitudes reached $0.02249$ for potentiation and $0.04262$ for depression, giving $A_{\mathrm{sym}}=-0.309$. Local reversals affected $45.2\%$ of potentiation increments and $50.8\%$ of depression increments. Despite only a small difference between its mean initial and final states, the depression branch covered a normalized range of approximately $0.230$. The combination of pronounced point-to-point modulation and limited net displacement is also consistent with the poor performance of the common smooth parametric law for this branch.

Sahu showed the coarsest potentiation increments and comparatively large depression increments [Figure~\ref{fig:device-characterization}(F)]. Their median values were $0.03651$ and $0.02546$, respectively, with $A_{\mathrm{sym}}=0.178$. Local direction violations occurred for $16.7\%$ of potentiation increments and $12.5\%$ of depression increments. Across all six models, the median programming-step magnitude covered approximately a 63-fold range, from the Pan depression branch to the Ioannou depression branch.

Native memory behavior introduced a second source of device diversity [Figure~\ref{fig:device-characterization}(G)]. For Luo, the post-stimulation relaxation reported in Figure~2 of Luo \textit{et al.} \cite{luo2022organic} was represented by the fitted stretched-exponential model. The resulting parameters were $\tau=1.32$~s, $\beta=0.794$, and $\rho=0.0512$. Over the available $29.1$~s observation interval, the locally normalized state changed by approximately $-43.7\%$. The reference level follows the source-specific relaxation normalization and is not interpreted as an independently measured equilibrium conductance.

Pan exhibited a different memory regime. The five post-stimulation trajectories in Figure~2g of Pan \textit{et al.} \cite{pan2024tio2wo}, corresponding to 20, 50, 100, 150, and 200 preceding stimulation pulses, had a median observed relative change of $-72.7\%$. Individual trajectories ranged from $-76.5\%$ to $-47.4\%$. Median observed-change crossing times were $t_{50}=1.40$~s and $t_{1/e}=2.24$~s, with respective ranges of $0.867$--$2.34$~s and $1.15$--$4.10$~s. These values describe the observed start-to-end transition and are not treated as fitted physical time constants. The relaxation curves were kept in the native scale of Figure~2g because the source study does not provide an experimentally established absolute mapping to the programming quantity in Figure~2h of Pan \textit{et al.} \cite{pan2024tio2wo}.

Memory dynamics were considerably slower and state dependent for Watanabe. The measurements from Figure~4a of Watanabe \textit{et al.} \cite{watanabe2024pedot} were retained as 12 electrolyte- and initial-state-specific trajectories [Figure~\ref{fig:device-characterization}(G)]. Their median relative observed change was $-2.49\%$, while individual responses ranged from a $20.7\%$ decrease to a $27.7\%$ increase. Among trajectories reaching the relevant observed-change thresholds, median $t_{50}$ was approximately $806$~s, with a range of $278$--$2.91\times10^{3}$~s. Median $t_{1/e}$ was approximately $1.25\times10^{3}$~s and ranged from $374$ to $2.96\times10^{3}$~s. The electrolyte-specific responses were not collapsed into a single effective relaxation curve.

Ioannou also showed pronounced state dependence. Relaxation measurements from Figure~4 and the Supplementary Information of Ioannou \textit{et al.} \cite{ioannou2020libased} were represented by 15 empirical trajectories grouped into post-depression, post-potentiation, and pulse-width-dependent source families [Figure~\ref{fig:device-characterization}(G)]. Median relative observed change was $-31.1\%$, with individual trajectories ranging from $-62.3\%$ to $+33.5\%$. Median observed $t_{50}$ was $3.96$~s, spanning approximately $0.12$ to $42.5$~s. The simple-exponential timescale of approximately $30$~s reported by Ioannou \textit{et al.} \cite{ioannou2020libased} was retained only as a source-derived descriptor and was not used as a common fitted relaxation constant.

Zeng and Sahu occupied the persistent end of the memory comparison [Figure~\ref{fig:device-characterization}(G)]. No reliably resolved monotonic decay was present over the retention interval reported in Figure~2c of Zeng \textit{et al.} \cite{zeng2023organic}. The measurements in Figures~2e and 3b of Sahu \textit{et al.} \cite{sahu2023hfox} likewise did not permit identification of a finite relaxation time. Native memory for both devices was represented by the experimentally supported lower bound $T_{\mathrm{ret}}\geq10^{4}$~s, without assigning finite $\tau$, $\beta$, or $\rho$.

Figure~\ref{fig:device-characterization}(H) collects the contrast in programming granularity, directional asymmetry, and native memory. Pan combines the smallest characteristic programming increments with strong potentiation--depression asymmetry, while Ioannou and Sahu operate with substantially coarser normalized updates. The memory characteristics are equally heterogeneous. Luo shows a resolved seconds-scale transient, Watanabe much slower electrolyte-dependent responses, Pan and Ioannou strong history- or state-dependent relaxation, and Zeng and Sahu persistence over the experimentally supported $10^{4}$~s interval. Complete branch-level programming statistics and native-memory descriptors are reported in Supplementary Tables~S3 and~S4.

\subsection{Device constraints affect stationary sEMG decoding}
\label{sec:results_stationary}

\begin{figure*}[ht!]
    \centering
    \includegraphics[width=\textwidth]{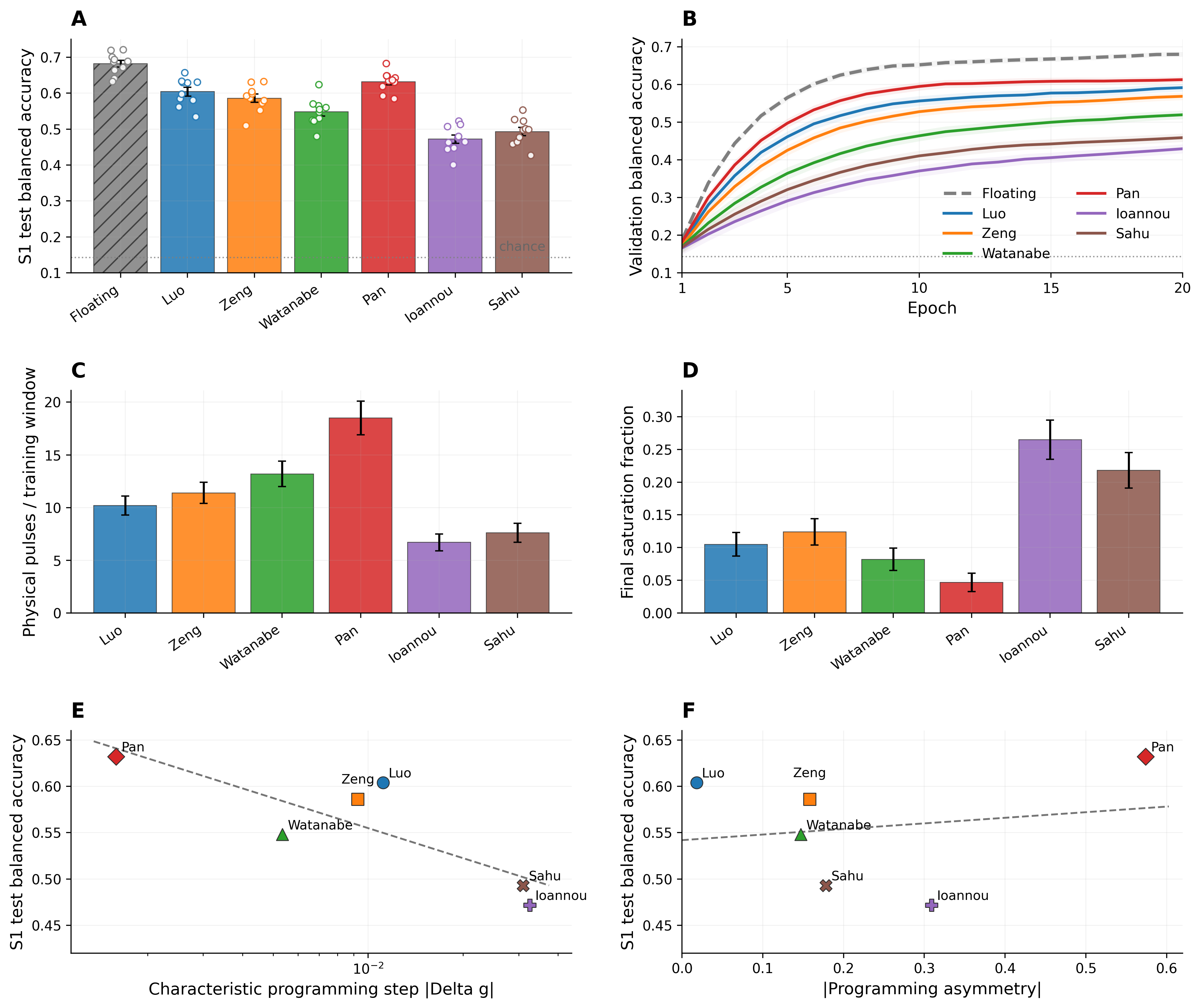}
    \caption{
Stationary Session-1 decoding under literature-constrained memristive
programming.
(A) Test balanced accuracy of the floating-point reference and the six
memristive implementations after Session-1 training; individual markers denote
participant-level values and the horizontal dotted line indicates chance
performance, $1/7$.
(B) Validation balanced accuracy across the 20 training epochs for the
floating-point and device-constrained networks.
(C) Mean physical programming burden per training window.
(D) Fraction of physical device states occupying the accessible programming
boundaries after Session-1 training.
(E) Session-1 test balanced accuracy as a function of the characteristic
normalized programming-step magnitude.
(F) Session-1 test balanced accuracy as a function of the absolute
potentiation--depression asymmetry index. Dashed lines in (E,F) are descriptive
trend lines and are not used for model fitting or statistical inference.
}
\label{fig:stationary}
\end{figure*}

We first tested whether all six literature-constrained synaptic models could support stationary sEMG decoding under the same network architecture and training protocol. This analysis used Session~1 only, with native relaxation disabled. Under this condition, differences in decoding performance reflected the programming characteristics of the synaptic models rather than temporal memory effects. Figure~\ref{fig:stationary} compares stationary performance with the corresponding programming characteristics. The floating-point implementation was retained as the reference.

The floating-point reference reached a mean test balanced accuracy of approximately $0.68$ [Figure~\ref{fig:stationary}(A)]. All memristive implementations remained below this level but above chance, $1/7 \approx 0.143$. Pan achieved the highest device-level accuracy at approximately $0.63$. Luo and Zeng followed at about $0.61$ and $0.59$, respectively, while Watanabe reached approximately $0.55$. Sahu and Ioannou gave lower values of about $0.49$ and $0.47$. The spread between the highest- and lowest-performing memristive models was approximately 0.16 in balanced accuracy.

These differences developed throughout training rather than appearing only at the final evaluation [Figure~\ref{fig:stationary}(B)]. The floating-point network improved rapidly during the first epochs and approached a validation balanced accuracy of about $0.68$. Pan followed most closely and reached approximately $0.61$--$0.62$. Luo and Zeng stabilized near $0.59$ and $0.57$--$0.58$. Watanabe converged more slowly toward approximately $0.53$--$0.54$, whereas Sahu and Ioannou remained lower throughout training, reaching roughly $0.47$ and $0.44$--$0.45$. Because the learning protocol was identical, this divergence reflects the conductance transitions imposed by the experimentally derived programming trajectories.

Programming demand also varied substantially across devices. Pan required approximately $18.5$ physical programming pulses per training window, the largest burden among the six models. Ioannou required about $6.7$ and Sahu about $7.7$ pulses/window. Watanabe occupied an intermediate position at approximately $13.2$ pulses/window, while Zeng and Luo required about $11.3$ and $10.2$, respectively [Figure~\ref{fig:stationary}(C)]. The $6.7$--$18.5$ pulses/window range amounts to an almost threefold difference in programming activity. Pan combined comparatively high decoding accuracy with the largest pulse burden, whereas the lower pulse counts of Ioannou and Sahu coincided with lower stationary performance.

Final occupation of the accessible device-state ranges followed a different pattern. Pan had the smallest saturation fraction, approximately $0.05$ [Figure~\ref{fig:stationary}(D)]. Watanabe remained near $0.08$, Luo reached approximately $0.11$, and Zeng approximately $0.13$. Sahu and Ioannou showed the largest fractions at about $0.22$ and $0.27$. This distinction becomes relevant in the subsequent continual-learning experiments because synapses already located at an accessible boundary after Session~1 have less conductance range available for later adaptation.

Figure~\ref{fig:stationary}(E) relates stationary accuracy to programming granularity. Pan combined the smallest characteristic single-step conductance change, $\widetilde{|\Delta g|}\approx1.6\times10^{-3}$, with the highest device-level balanced accuracy of approximately $0.63$. Watanabe, with $\widetilde{|\Delta g|}\approx5.3\times10^{-3}$, reached approximately $0.55$. Zeng and Luo had characteristic steps of about $9.3\times10^{-3}$ and $1.1\times10^{-2}$ and balanced accuracies near $0.59$ and $0.61$. Sahu and Ioannou occupied the opposite end of the range, with steps of approximately $3.1\times10^{-2}$ and $3.3\times10^{-2}$ and accuracies of about $0.49$ and $0.47$. Across these six points, stationary accuracy tended to decrease as characteristic step size increased, although Luo, Zeng, and Watanabe show that step size alone is not sufficient to determine the result.

Programming asymmetry had an even less regular relationship with decoding accuracy [Figure~\ref{fig:stationary}(F)]. Luo combined an almost symmetric response, $|A_{\mathrm{sym}}|\approx0.02$, with a balanced accuracy of approximately $0.61$. Zeng and Watanabe both had $|A_{\mathrm{sym}}|\approx0.15$, yet their accuracies differed by several percentage points. Sahu and Ioannou showed larger asymmetry magnitudes of approximately $0.18$ and $0.31$ together with lower stationary performance. Pan breaks any simple monotonic interpretation most clearly: despite the largest asymmetry, $|A_{\mathrm{sym}}|\approx0.57$, it also achieved the highest balanced accuracy among the memristive models. 

The six implementations entered the continual-learning experiments from different operating points even before temporal relaxation was enabled. Session~1 balanced accuracy ranged from approximately $0.47$ to $0.63$, programming demand from $6.7$ to $18.5$ physical pulses per training window, and final saturation from approximately $0.05$ to $0.27$. All six models could acquire the stationary task, but they did so under substantially different physical programming constraints.

\subsection{Inter-session shifts separate adaptation from retention}
\label{sec:results_sessions}

\begin{figure*}[ht!]
    \centering
    \includegraphics[width=\textwidth]{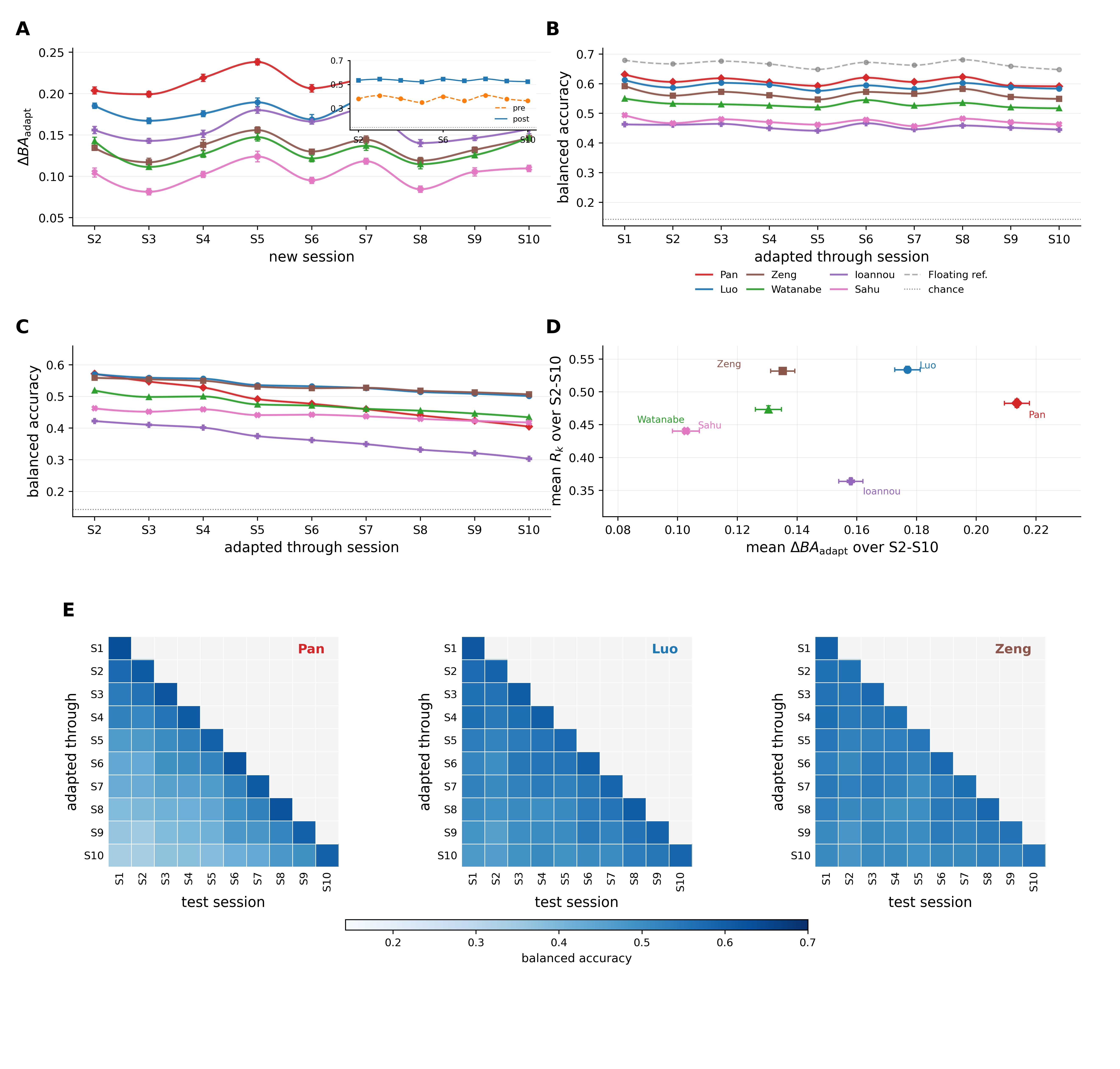}
    \caption{
Adaptation and retention during sequential sEMG sessions.
(A) Mean change in balanced accuracy between pre-adaptation and
post-adaptation evaluation for Sessions $S_2$--$S_{10}$; the inset illustrates
the corresponding pre- and post-adaptation levels.
(B) Balanced accuracy on the current-session test set after sequential
adaptation through $S_1$--$S_{10}$, with the floating-point reference and
chance level shown for comparison.
(C) Mean balanced accuracy across previously encountered test sessions after
adaptation through each new session.
(D) Device-level relation between mean adaptation gain and mean retained
balanced accuracy across Sessions $S_2$--$S_{10}$.
(E) Lower-triangular $BA_{k,\ell}$ matrices for Pan, Luo, and Zeng, showing
balanced accuracy on each previously encountered test session $\ell$ after
adaptation through session $k$. The undefined upper-triangular region
corresponds to future sessions that had not yet been encountered.
}
\label{fig:session_dynamics}
\end{figure*}

Starting from the Session~1 endpoint, each network was adapted sequentially to Sessions~2--10 under the same continual-learning protocol. Performance was measured immediately before and after adaptation to each new session, while the test sets of previously encountered sessions were retained for read-only evaluation. Figure~\ref{fig:session_dynamics} separates recovery on the current session from preservation of earlier ones.

Every session transition produced a measurable increase in balanced accuracy after adaptation, but the magnitude varied considerably between devices [Figure~\ref{fig:session_dynamics}(A)]. Averaged across Sessions~2--10, Pan gave the largest adaptation gain, $\overline{\Delta BA}_{\mathrm{adapt}}=0.214$. Luo followed at $0.177$, and Ioannou at $0.158$. Zeng, Watanabe, and Sahu showed smaller mean gains of $0.135$, $0.130$, and $0.103$, respectively. The exact ordering changed at individual session transitions. Around S5, for example, Pan reached approximately $0.239$, Luo $0.190$, and Ioannou $0.180$, while several models showed weaker recovery around S3 and S8. 

Current-session decoding remained functional throughout the complete S1--S10 sequence [Figure~\ref{fig:session_dynamics}(B)]. Pan decreased from $BA_{1,1}=0.628$ in S1 to approximately $BA_{10,10}=0.590$ after S10. Luo changed from $0.611$ to $0.570$, and Zeng from $0.590$ to $0.550$. Watanabe moved from approximately $0.552$ to $0.518$. Sahu changed from $0.489$ to $0.458$, while Ioannou decreased from $0.462$ to approximately $0.447$. The floating-point reference remained between approximately $0.64$ and $0.68$ over the same sequence. None of the device models approached the chance level of $1/7$. Recurrent increases at several sessions also show that adaptation to the current sEMG distribution continued throughout the sequence.

The ordering changed when earlier sessions were considered. Figure~\ref{fig:session_dynamics}(C) reports the mean balanced accuracy over all previously encountered test sessions after adaptation through each $S_k$. Zeng showed the strongest overall retention profile, remaining close to $R_2=0.56$ after the first session transition and reaching approximately $R_{10}=0.50$ after S10. Luo followed a similar trajectory, decreasing from approximately $0.57$ to $0.50$. Pan began at a comparable level, about $R_2=0.57$, but declined more strongly to approximately $R_{10}=0.40$. Watanabe decreased from approximately $0.51$ to $0.43$, and Sahu from about $0.46$ to $0.42$. Ioannou showed the largest loss of previously acquired information, with mean retention falling from approximately $0.42$ after S2 to $0.30$ after S10.

The two-dimensional comparison in Figure~\ref{fig:session_dynamics}(D) makes this separation more explicit. Pan had the largest mean adaptation gain, $0.214$, but its mean retention over S2--S10 was only about $0.48$. Zeng showed the reverse pattern, with a lower adaptation gain of $0.135$ and mean retention of approximately $0.535$. Luo combined an adaptation gain of $0.177$ with retention near $0.53$. Watanabe occupied an intermediate position at approximately $(0.130,\,0.474)$. Sahu showed limited adaptation, $0.103$, and mean retention of about $0.44$. Ioannou combined an adaptation gain of $0.158$ with the lowest mean retention, approximately $0.37$. 

The lower-triangular $BA_{k,l}$ matrices in Figure~\ref{fig:session_dynamics}(E) show the same distinction for individual session pairs. Pan retains comparatively high diagonal values across S1--S10, while performance on earlier sessions decreases as additional sessions are learned. Luo has a more uniform lower triangle, with relatively high current-session accuracy and a slower decline in earlier-session performance. Zeng preserves the lower-triangular structure most strongly among the three examples, although its diagonal values remain below those of Pan. The upper triangle is undefined because future-session test sets were not evaluated before the corresponding adaptation state was reached.

Pan responds strongly to each new input distribution but loses more of its earlier performance. Zeng changes less at each adaptation step and preserves earlier sessions more effectively. Luo maintains comparatively high values of both quantities. These differences are quantified below using the continual-learning metrics and the native memory characteristics of the device models.

\subsection{Device dynamics determine the stability--plasticity trade-off}
\label{sec:results_stability}

\begin{figure*}[ht!]
    \centering
    \includegraphics[width=\textwidth]{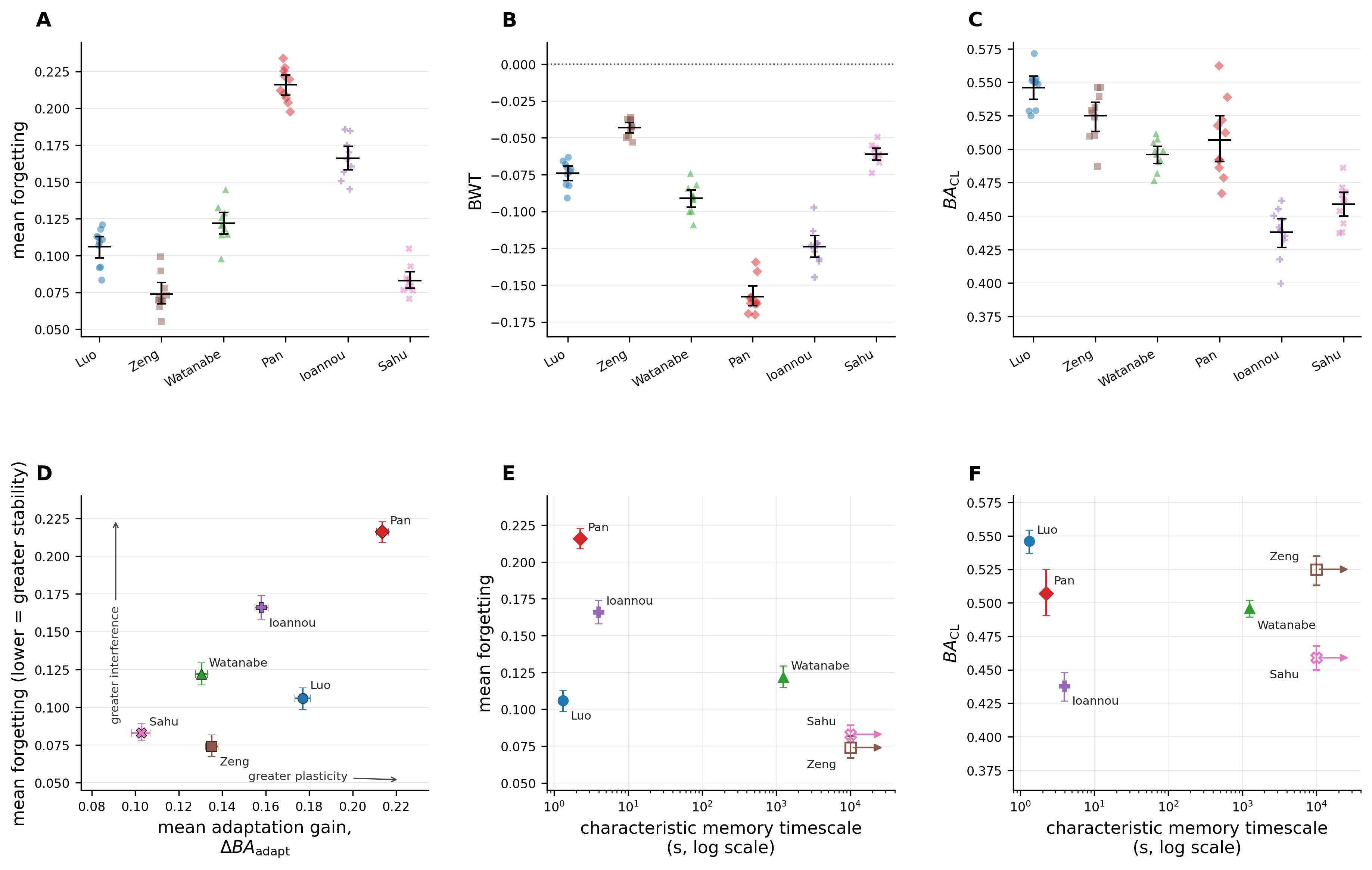}
    \caption{
Device-dependent stability--plasticity characteristics during continual sEMG
decoding.
(A) Mean forgetting across the six literature-constrained synaptic models.
(B) Backward transfer (BWT), with zero indicating the absence of net backward
interference.
(C) Continual-learning balanced accuracy $BA_{\mathrm{CL}}$.
(D) Stability--plasticity plane relating mean adaptation gain
$\Delta BA_{\mathrm{adapt}}$ to mean forgetting; lower forgetting indicates
greater stability, whereas increasing adaptation gain indicates greater
plasticity.
(E) Mean forgetting plotted against the source-specific characteristic memory
descriptor of each device.
(F) Continual-learning balanced accuracy plotted against the same memory
descriptors. Horizontal arrows for Zeng and Sahu indicate experimentally
supported lower bounds on retention rather than resolved finite relaxation
times.
}
\label{fig:stability_plasticity}
\end{figure*}

The session-level differences were quantified using the continual-learning metrics defined in Section~\ref{sec:evaluation_statistics}. All six memristive implementations remained capable of acquiring the current session, but the interference introduced into previously learned sessions varied substantially. Figure~\ref{fig:stability_plasticity} compares forgetting, backward transfer, and aggregate continual-learning performance across the six models.

Mean forgetting ranged from $0.074$ to $0.214$ [Figure~\ref{fig:stability_plasticity}(A)]. Zeng showed the lowest value, $0.074$, followed by Sahu at $0.083$ and Luo at $0.106$. Watanabe reached $0.122$. Larger losses occurred for Ioannou and Pan, with forgetting values of $0.165$ and $0.214$. This ordering differed from the stationary Session~1 result, where Pan had the highest device-level decoding accuracy.

Backward transfer followed the same pattern of sequential interference [Figure~\ref{fig:stability_plasticity}(B)]. BWT was negative for all six models, indicating that later adaptation reduced performance on earlier sessions. Zeng remained closest to zero at $\mathrm{BWT}=-0.042$, followed by Sahu at $-0.061$ and Luo at $-0.075$. Watanabe reached $-0.091$. Ioannou and Pan showed the strongest interference, with values of $-0.123$ and $-0.158$. Pan preserved its strong response to the current input distribution at the cost of a larger loss of previous representations.

Aggregate continual-learning performance produced another ordering [Figure~\ref{fig:stability_plasticity}(C)]. Luo achieved the highest $BA_{\mathrm{CL}}$ among the memristive models, $0.547$, despite not having the largest adaptation gain. Zeng reached $0.526$, Pan $0.507$, and Watanabe $0.495$. Sahu and Ioannou were lower at $0.459$ and $0.439$. The stationary advantage of Pan was reduced once preservation of earlier sessions became part of the evaluation.

The stability--plasticity plane in Figure~\ref{fig:stability_plasticity}(D) shows where this reordering originates. Pan combined the highest mean adaptation gain, $\overline{\Delta BA}_{\mathrm{adapt}}=0.214$, with the highest forgetting, also $0.214$. Zeng occupied the more stable end of the plane, with adaptation gain $0.135$ and forgetting $0.074$. Sahu also showed low forgetting, $0.083$, but its adaptation gain of $0.103$ and lower current-session performance limited $BA_{\mathrm{CL}}$. Luo occupied a different region at $(0.177,\,0.106)$ and produced the highest overall continual accuracy. Watanabe remained intermediate at approximately $(0.130,\,0.122)$, while Ioannou combined adaptation gain $0.158$ with forgetting $0.165$. 

Their native memory characteristics provide one physical axis for this separation. Pan, with an observed $t_{1/e}$ of approximately $2.24$~s, had the largest forgetting, $0.214$ [Figure~\ref{fig:stability_plasticity}(E)]. Ioannou, whose median observed $t_{50}$ was approximately $3.96$~s, reached $0.165$. Luo did not follow a simple monotonic pattern. Its fitted relaxation timescale was approximately $1.32$~s, yet mean forgetting was only $0.106$. Watanabe occupied a much longer temporal range, with median observed $t_{1/e}$ of approximately $1.25\times10^3$~s and forgetting of $0.122$. Zeng and Sahu showed no resolved decay over the reported retention interval, represented as lower bounds of $\geq10^{4}$~s, and their forgetting values were correspondingly low at $0.074$ and $0.083$.

Memory persistence alone still failed to determine continual-learning performance [Figure~\ref{fig:stability_plasticity}(F)]. Zeng and Sahu both had retention bounds of $\geq10^{4}$~s, but their $BA_{\mathrm{CL}}$ values differed, $0.526$ versus $0.459$. The shorter-memory models also separated. Luo reached $BA_{\mathrm{CL}}=0.547$ at approximately $1.32$~s, Pan $0.507$ at $2.24$~s, and Ioannou only $0.439$ at $3.96$~s. Watanabe, at approximately $1.25\times10^{3}$~s, produced an intermediate value of $0.495$. Longer memory could reduce forgetting, but its contribution to continual performance depended on more than lifetime alone. The controlled experiment below compares synaptic memory directly with the rate of input drift.

\subsection{Synaptic memory and input drift reveal a common timescale dependence}
\label{sec:results_timescale}

\begin{figure*}[ht!]
    \centering
    \includegraphics[width=\textwidth]{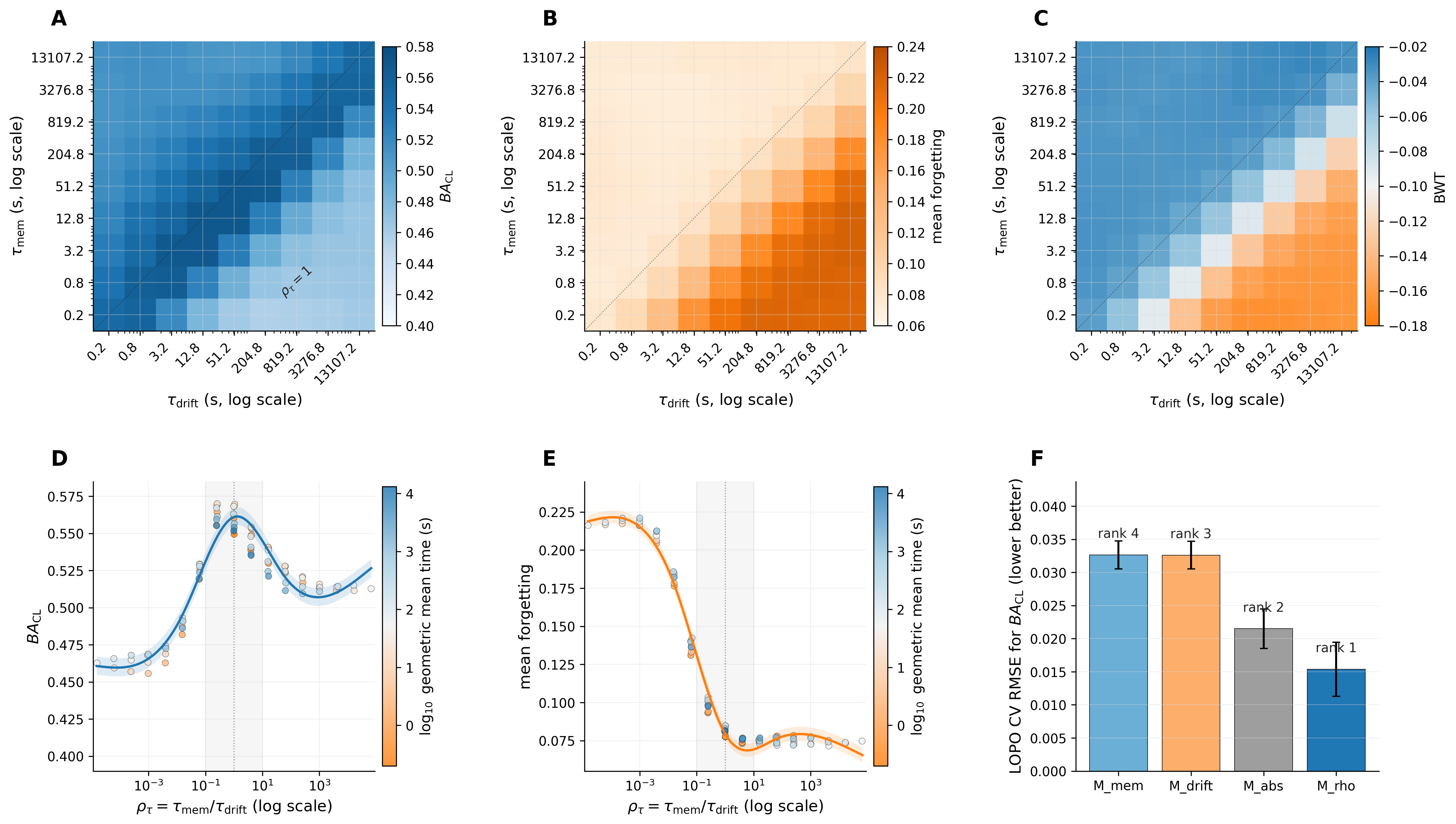}
    \caption{
Controlled dependence of continual-learning performance on synaptic-memory and
input-drift timescales.
(A) Continual-learning balanced accuracy $BA_{\mathrm{CL}}$ across the
$(\tau_{\mathrm{mem}},\tau_{\mathrm{drift}})$ plane; the diagonal indicates
$\rho_\tau=1$.
(B) Mean forgetting over the same controlled timescale grid.
(C) Backward transfer over the controlled timescale grid.
(D) Collapse of $BA_{\mathrm{CL}}$ as a function of the relative timescale
$\rho_\tau=\tau_{\mathrm{mem}}/\tau_{\mathrm{drift}}$; point color denotes the
geometric mean of the two absolute timescales and the solid curve is a
descriptive smooth dependence.
(E) Corresponding dependence of mean forgetting on $\rho_\tau$.
(F) Leave-one-participant-out cross-validation RMSE for models based on
$\tau_{\mathrm{mem}}$, $\tau_{\mathrm{drift}}$, both absolute timescales, and
the relative timescale $\rho_\tau$. Lower RMSE indicates better predictive
performance.
}
\label{fig:timescale_matching}
\end{figure*}

The native-device comparison left memory persistence unable to account for continual-learning performance by itself. To isolate the temporal contribution, synaptic memory and input drift were varied independently of the native relaxation characteristics. Both $\tau_{\mathrm{mem}}$ and $\tau_{\mathrm{drift}}$ covered the same nine-point logarithmic grid from $0.2$ to $13107.2$~s, giving $81$ controlled combinations. This construction places the two timescales on the same numerical grid and removes the ambiguity introduced by the different native memory descriptors. The resulting dependence is shown in Figure~\ref{fig:timescale_matching}.

The $BA_{\mathrm{CL}}$ surface did not align with either absolute timescale independently [Figure~\ref{fig:timescale_matching}(A)]. Instead, the high-performance region extended diagonally through the $(\tau_{\mathrm{mem}},\tau_{\mathrm{drift}})$ plane. The lowest values, approximately $BA_{\mathrm{CL}}=0.458$, appeared when synaptic memory was short relative to the drift timescale. Performance increased toward a broad ridge and reached $BA_{\mathrm{CL}}=0.565$ as the two temporal scales became comparable. It then declined as synaptic persistence increasingly exceeded the rate of input change. The ridge was not confined to $\tau_{\mathrm{mem}}=\tau_{\mathrm{drift}}$, indicating tolerance to moderate mismatch.

Forgetting produced a complementary surface [Figure~\ref{fig:timescale_matching}(B)]. Mean forgetting approached $0.218$ when memory was short and drift was slow. Increasing $\tau_{\mathrm{mem}}$ relative to $\tau_{\mathrm{drift}}$ reduced this loss, first below $0.10$ over a broad region and eventually to a minimum near $0.070$. Beyond that region, further increases in memory timescale produced little additional reduction.

Backward transfer was organized in the same way [Figure~\ref{fig:timescale_matching}(C)]. The most negative value, approximately $\mathrm{BWT}=-0.176$, occurred when synaptic memory decayed rapidly relative to the evolving input. BWT moved toward zero as relative memory lifetime increased and reached approximately $-0.025$ in the persistent-memory region. The BWT and forgetting surfaces identify the same loss of earlier information at short relative memory lifetime. The $BA_{\mathrm{CL}}$ surface, however, shows that the condition minimizing forgetting is not the condition with the highest overall continual performance.

Expressing the $81$ combinations through Equation~(\ref{eq:timescale_ratio}) revealed the diagonal structure more directly [Figure~\ref{fig:timescale_matching}(D)]. At $\rho_{\tau}\ll1$, $BA_{\mathrm{CL}}$ remained around $0.460$--$0.470$. It then rose sharply, reaching a maximum of approximately $0.565$ for $\rho_{\tau}$ slightly below unity. At $\rho_{\tau}=1$, performance remained high at approximately $0.558$. Larger ratios reduced $BA_{\mathrm{CL}}$, which approached approximately $0.510$ for $\rho_{\tau}$ between $10^{2}$ and $10^{4}$. A modest increase reappeared at the most extreme ratios, although performance remained below the intermediate-ratio maximum. Points with similar $\rho_{\tau}$ retained some spread, so absolute timescales still had a secondary effect and the collapse was not exact.

The dependence of forgetting on $\rho_{\tau}$ was more nearly monotonic [Figure~\ref{fig:timescale_matching}(E)]. Mean forgetting remained around $0.215$--$0.220$ for $\rho_{\tau}\lesssim10^{-2}$ and then dropped rapidly through the transition region. It reached approximately $0.098$ near $\rho_{\tau}=10^{-1}$ and approximately $0.081$ at $\rho_{\tau}=1$. The minimum, approximately $0.070$, occurred above unity. At still larger ratios, forgetting remained low, generally between $0.072$ and $0.080$. The ratio giving minimum forgetting extended further into the persistent regime than the ratio producing maximum $BA_{\mathrm{CL}}$.

This organization was also reflected in the predictive comparison [Figure~\ref{fig:timescale_matching}(F)]. Under leave-one-participant-out cross-validation, the model based on $\rho_{\tau}$ gave the lowest RMSE, $0.0154$. The additive model containing both absolute timescales, $M_{\mathrm{abs}}$, reached $0.0215$. Models using either absolute timescale alone performed worse, with RMSE values of $0.0328$ for $M_{\mathrm{mem}}$ and $0.0327$ for $M_{\mathrm{drift}}$. Relative to $M_{\mathrm{mem}}$, prediction error decreased by approximately $53.0\%$ for $M_{\rho}$; relative to $M_{\mathrm{drift}}$, the reduction was approximately $52.9\%$. The complete leave-one-participant-out comparison is reported in Supplementary Table~S5.

The controlled sweep also explains why native device behavior could not be organized by memory lifetime alone. Similar absolute retention can produce different continual-learning outcomes when the input changes on a different temporal scale. Conversely, markedly different absolute timescales can produce similar behavior when their ratio is maintained. This raises the question of whether two synaptic processes with different memory lifetimes can occupy a more favorable adaptation--retention regime when combined.

\subsection{Hybrid fast--persistent synapses alter the adaptation--retention trade-off}
\label{sec:results_hybrid}

\begin{figure*}[ht!]
    \centering
    \includegraphics[width=\textwidth]{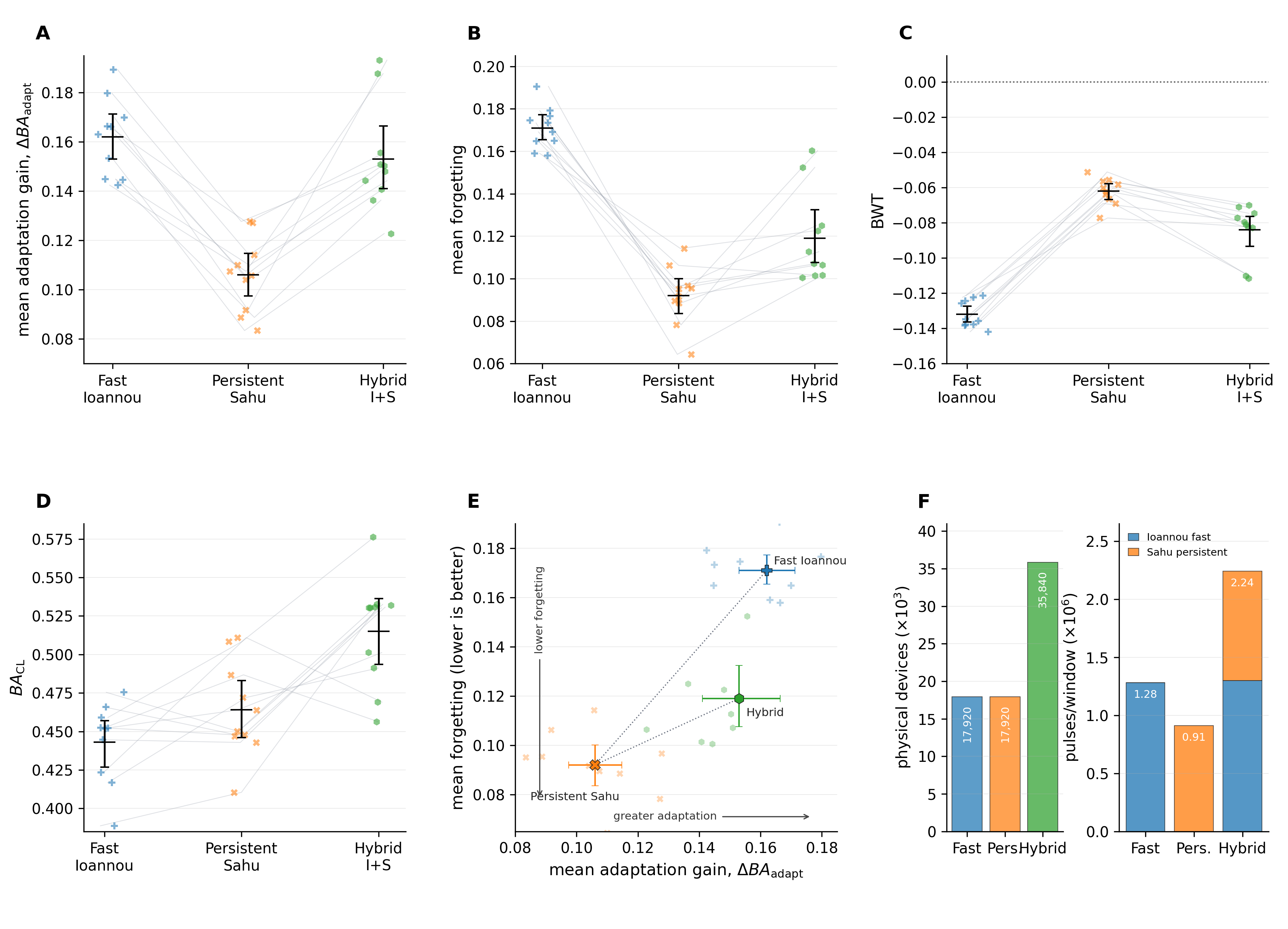}
   \caption{
Continual-learning performance of the hybrid fast--persistent synaptic
representation.
(A) Mean adaptation gain for the fast Ioannou, persistent Sahu, and hybrid
Ioannou--Sahu implementations.
(B) Mean forgetting of previously encountered sessions.
(C) Backward transfer.
(D) Continual-learning balanced accuracy $BA_{\mathrm{CL}}$.
In panels (A--D), participant-level values are shown together with paired
trajectories across the three synaptic implementations.
(E) Adaptation--forgetting plane illustrating the operating regions of the fast,
persistent, and hybrid synapses; increasing adaptation corresponds to greater
plasticity and decreasing forgetting to greater stability.
(F) Hardware cost of the three implementations, expressed as the number of
physical memristive elements and cumulative programming burden. The hybrid
contains separate Ioannou and Sahu differential branches, doubling the number
of physical elements per logical synapse relative to either single-device
implementation.
}
\label{fig:hybrid}
\end{figure*}

The controlled analysis showed reduced continual decoding when synaptic dynamics were either too transient or too persistent relative to input evolution. We combined the fast Ioannou branch with the persistent Sahu branch, using equal contributions to the effective synaptic weight and no additional consolidation rule. Both branches received the same learning signal and evolved independently according to their device-specific dynamics. Figure~\ref{fig:hybrid} compares this hybrid with the two corresponding single-device implementations.

The two individual components already differed in their response to new sessions [Figure~\ref{fig:hybrid}(A)]. Ioannou produced a mean adaptation gain of $\overline{\Delta BA}_{\mathrm{adapt}}=0.158$, while Sahu reached $0.103$. The hybrid retained most of the fast component's adaptive response, with $\overline{\Delta BA}_{\mathrm{adapt}}=0.151$. This was $0.007$ below Ioannou and $0.048$ above Sahu. Participant-level trajectories followed the same overall pattern, although the distributions overlapped and the magnitude of the hybrid response varied between participants.

The ordering reversed when earlier sessions were considered. Mean forgetting was $0.165$ for Ioannou and $0.083$ for Sahu [Figure~\ref{fig:hybrid}(B)]. The hybrid reached $0.119$, reducing forgetting by $0.046$ relative to Ioannou while remaining $0.036$ above Sahu. It did not reproduce the minimum forgetting of the persistent device, but moved away from the high-forgetting behavior of the fast component while retaining much of its adaptation capability.

Backward transfer showed the same redistribution [Figure~\ref{fig:hybrid}(C)]. Ioannou had the most negative value, $\mathrm{BWT}=-0.123$, and Sahu remained closer to zero at $-0.061$. The hybrid reached $-0.083$, an improvement of $0.040$ relative to Ioannou while remaining $0.022$ below Sahu. BWT stayed negative in all three cases, so later adaptation continued to interfere with earlier sessions even in the hybrid implementation.

These opposing changes altered the ordering of overall continual-learning performance [Figure~\ref{fig:hybrid}(D)]. Ioannou reached $BA_{\mathrm{CL}}=0.439$ and Sahu $0.459$. The hybrid increased this value to $0.516$, exceeding the fast and persistent implementations by $0.077$ and $0.057$, respectively. Participant-level values still overlapped substantially, with the hybrid ranging from approximately $0.456$ to $0.576$. The difference in mean continual accuracy reflected a shift in the paired participant responses rather than complete separation of the three distributions.

The adaptation--forgetting plane in Figure~\ref{fig:hybrid}(E) shows the corresponding change in operating point. Ioannou occupied $(0.158,\,0.165)$, the high-adaptation, high-forgetting region. Sahu occupied $(0.103,\,0.083)$, with lower adaptation and lower forgetting. The hybrid moved to $(0.151,\,0.119)$. It retained $95.6\%$ of the mean adaptation gain of Ioannou while reducing forgetting by $27.9\%$. Relative to Sahu, adaptation increased by $46.6\%$, accompanied by a $43.4\%$ increase in forgetting. 

This shift required additional hardware resources. A conventional differential synapse uses two memristive elements per logical connection. With $8960$ logical synapses in the $28\rightarrow256\rightarrow7$ network, either single-device implementation required $17\,920$ physical memristive elements [Figure~\ref{fig:hybrid}(F)]. Two differential branches per logical synapse increased the hybrid count to $35\,840$. Programming activity also rose. Ioannou required $1.28\times10^{6}$ programming pulses and Sahu $0.91\times10^{6}$, while the combined hybrid burden reached $2.24\times10^{6}$ pulses. The higher continual-learning performance was accompanied by a twofold increase in device count and a larger total programming burden. The corresponding adaptation, retention, continual-learning, and hardware burden metrics for the primary Ioannou--Sahu hybrid, together with the sensitivity analyses for Ioannou--Zeng, Pan--Sahu, and Pan--Zeng, are summarized in Supplementary Table~S6.


\section{Discussion}
\label{sec:discussion}

Across the six literature-constrained models, neither material class nor any single device descriptor was sufficient to anticipate continual-learning behavior. Programming granularity, directional asymmetry, accessible conductance range, relaxation dynamics, and retention all differed between devices, and these differences remained visible after the models were placed in the same SNN. The stationary experiment provides a clear example. Pan achieved the highest device-level Session-1 balanced accuracy despite having the strongest programming asymmetry. Ioannou and Sahu, by comparison, produced lower stationary accuracy while requiring fewer programming pulses. Update symmetry, step size, and pulse count describe only part of the behavior relevant to continual decoding.

A clearer separation emerged during sequential adaptation. Pan combined the largest mean adaptation gain, $\Delta BA_{\mathrm{adapt}}=0.214$, with the largest forgetting, $F=0.214$. Zeng adapted less strongly but showed the smallest forgetting, $F=0.074$. Luo occupied another part of this trade-off and reached the highest continual-learning balanced accuracy among the six single-device implementations, $BA_{\mathrm{CL}}=0.547$. Device selection for a continually adapting decoder differs from selection based only on how closely a memristor reproduces an ideal stationary weight update. Here, the behavior of the physical synapse also depends on how its dynamics relate to changes in the physiological input.

\subsection{Memory timescale and environmental non-stationarity}
\label{sec:discussion_timescale}

Persistence alone did not explain the native-device continual-learning results. Zeng and Sahu both had experimentally supported retention beyond $10^{4}$~s, yet their continual-learning accuracies were $BA_{\mathrm{CL}}=0.526$ and $0.459$, respectively. The seconds-scale devices were also non-monotonic. Luo, Pan, and Ioannou had source-specific memory descriptors within this temporal range but differed markedly in both forgetting and continual-learning performance. These native-device results necessarily combine programming dynamics, memory dynamics, and the particular experimental characterization available for each device. To separate these contributions, native free relaxation was replaced by the same controlled exponential law, allowing $\tau_{\mathrm{mem}}$ and $\tau_{\mathrm{drift}}$ to vary independently while the device-specific programming trajectories remained unchanged. The resulting $BA_{\mathrm{CL}}$ surface was not organized primarily by either absolute timescale. Its high-performance region instead formed a broad diagonal band in the $(\tau_{\mathrm{mem}},\tau_{\mathrm{drift}})$ plane. When the same conditions were expressed through Equation~(\ref{eq:timescale_ratio}), performance increased from approximately $0.46$ at very small $\rho_{\tau}$ to a maximum near $0.565$ for a ratio slightly below unity. At $\rho_{\tau}=1$, $BA_{\mathrm{CL}}$ remained high at approximately $0.558$.

At $\rho_{\tau}\ll1$, the synaptic state relaxes before the corresponding input distribution changes appreciably. Information acquired during adaptation is then lost on a shorter timescale than the process being tracked. Increasing $\tau_{\mathrm{mem}}$ reduces this mismatch and sharply lowers forgetting. At the opposite extreme, persistent synapses preserve previously acquired states, but further increases in persistence do not continue to improve $BA_{\mathrm{CL}}$. The highest continual performance appeared in the intermediate region, where memory remained sufficiently persistent to retain useful information while still changing on a timescale compatible with the evolving input.

A separate test came from the predictive models. The model based on $\rho_{\tau}$ reached a leave-one-participant-out RMSE of $0.0154$, compared with $0.0328$ for $\tau_{\mathrm{mem}}$ alone and $0.0327$ for $\tau_{\mathrm{drift}}$ alone. The additive model containing both absolute timescales produced an RMSE of $0.0215$. Within this predefined model comparison, the ratio captured structure that was not recovered by treating the two absolute timescales as separate additive predictors. The collapse was nevertheless incomplete. Residual variation at fixed $\rho_{\tau}$ shows that absolute timescales and device-specific programming dynamics still contribute to the response. For this reason, $\rho_{\tau}$ is better regarded as an organizing variable than as a complete one-parameter description of the system.

In hardware terms, a retention time considered by itself says little about whether that timescale is useful for a particular task. Short memory may be appropriate when the environment changes rapidly, whereas slower drift can require substantially longer persistence. The relevant synaptic timescale is linked to the temporal structure of the process being represented.

\subsection{Implications for continual myoelectric decoding}
\label{sec:discussion_semg}

sEMG decoding makes this issue particularly visible because the signal is intrinsically non-stationary. Electrode position, skin contact, contraction intensity, fatigue, and recruitment patterns can change the multichannel distribution between recording sessions even when the intended movement remains the same. Performance measured only within one session describes a narrower problem than repeated use of the interface.

Sequential adaptation further separated current-session recovery from retention of earlier sessions. Pan recovered the largest amount of current-session performance after adaptation but showed the strongest forgetting. Zeng exhibited smaller adaptation gains and stronger retention. Luo maximized neither quantity individually, yet it achieved the highest $BA_{\mathrm{CL}}$ among the single-device models. Differences that were less apparent in Session~1 became visible after repeated adaptation.

The session-resolved $BA_{k,\ell}$ matrices expose this history directly. High values along the diagonal indicate that the decoder can still learn the current session. Changes away from the diagonal show how later adaptation affects representations acquired earlier. Pan, Luo, and Zeng produced different lower-triangular patterns even when their current-session performance was comparatively similar. Pooling the ten DB6 sessions and randomly shuffling them before training would remove this temporal information.

Controlled electrode displacement served a narrower purpose. It is not intended as a complete model of physiological sEMG drift. Instead, it provides a reproducible perturbation whose characteristic timescale can be varied independently of synaptic memory. The natural DB6 sequence and the controlled drift experiment address different parts of the problem. DB6 provides recorded session-to-session variability; controlled displacement provides the parameter space needed to examine the dependence on $\rho_{\tau}$.

For neuromorphic sEMG decoding, this means that sequential evaluation can reveal behavior that stationary benchmarks cannot. A device with high single-session accuracy may move to an unfavorable operating region once repeated adaptation begins. Another device may sacrifice some stationary accuracy while retaining information more effectively over longer use. These differences become visible only when acquisition order is preserved.

\subsection{Hybrid fast and persistent memory}
\label{sec:discussion_hybrid}

The hybrid experiment examined whether combining two physical memory dynamics could shift the stability--plasticity relation observed for the single-device models. Ioannou and Sahu were selected because their native implementations occupied different regions of this comparison. Ioannou adapted relatively strongly but showed substantial forgetting, whereas Sahu was more persistent and less adaptive.

Combining the two branches changed this balance without adding a software consolidation rule, branch-specific learning rate, or adaptive routing mechanism. Mean adaptation gain remained at $0.151$, close to the $0.158$ obtained with Ioannou alone, while forgetting decreased from $0.165$ to $0.119$. Backward transfer changed from $-0.123$ for Ioannou to $-0.083$ for the hybrid. These changes raised $BA_{\mathrm{CL}}$ to $0.516$, compared with $0.439$ for Ioannou and $0.459$ for Sahu.

The hybrid did more than reproduce an intermediate continual-learning accuracy between its two components. Its $BA_{\mathrm{CL}}$ exceeded both single-device values. The fast branch retained much of the responsiveness to the changing input, while the persistent branch reduced the loss of previously acquired information. 

However, there is a direct hardware cost. A conventional differential synapse uses two physical memristive elements per logical connection. In the $28\rightarrow256\rightarrow7$ network, this corresponds to $17\,920$ devices. The two-branch hybrid requires $35\,840$. Its cumulative programming burden also increased to $2.24\times10^{6}$ pulses, compared with $1.28\times10^{6}$ for Ioannou and $0.91\times10^{6}$ for Sahu. The hybrid represents a performance--resource trade-off. It shows that combining physical timescales can improve continual decoding under the present protocol, without implying that duplicating every synaptic element is necessarily the preferred hardware implementation.

\subsection{Limitations}
\label{sec:limitations}

The six device models were reconstructed from experiments performed by different laboratories with different material stacks, geometries, pulse protocols, readout conditions, and reporting conventions. Device-local normalization was required to map these measurements to a common logical synaptic range, but it cannot remove this underlying heterogeneity. Cross-device differences should be interpreted as comparisons between literature-constrained models, not as controlled measurements of material classes.

Machine-readable source measurements were not available for every device. In those cases, the published curves were reconstructed by axis-calibrated digitization. This preserves the reported trajectories more closely than replacing them with a generic analytic memristor model, although finite digitization uncertainty remains. The empirical PCHIP representation also retains local reversals in the source curves. Such fluctuations may contain contributions from both device physics and measurement noise, but removing them would impose a smoother response than the available experiment supports.

Native memory characterization also differs between sources. Luo supports a fitted relaxation law. Watanabe and Ioannou provide multiple state- or condition-dependent trajectories, Pan provides history-dependent post-stimulation relaxation, and Zeng and Sahu provide retention lower bounds without a resolved finite decay. The native-device analysis uses source-specific descriptors rather than treating these quantities as equivalent physical time constants. The controlled $\tau_{\mathrm{mem}}$ experiment avoids this ambiguity when testing the timescale relation.

The paper is based on NinaPro DB6. The quantitative optimum in $\rho_{\tau}$ cannot be assumed to remain unchanged for other sEMG datasets, electrode layouts, acquisition systems, or clinical populations. Ten participants provide repeated multi-session recordings for testing the proposed mechanism, but they do not support a universal numerical optimum. Within the present protocol, the broader result is the dependence on relative memory and drift timescales and the lower prediction error obtained with $M_{\rho}$.

The network was simulated rather than fabricated. Device-constrained programming trajectories, differential weight representation, relaxation, pulse counts, and physical element counts were included, whereas array-level parasitic resistance, line capacitance, write disturb, finite peripheral precision, transistor mismatch, temperature dependence, device ageing, and endurance-related failures were not. Programming burden should consequently be interpreted as relative write activity rather than a direct prediction of device lifetime. Cross-device write energy was also excluded because the source studies do not provide a sufficiently uniform set of current and pulse-energy measurements.

\subsection{Future directions and hardware validation}
\label{sec:discussion_future}

A direct experimental test would require both timescales in the ratio to come from measured physical processes rather than controlled numerical parameters. On the signal side, characteristic drift times could be estimated from longitudinal sEMG recordings under defined perturbations such as electrode removal and replacement, controlled displacement, fatigue, or changes in contraction intensity. On the device side, synaptic relaxation would need to be measured and tuned under the same programming conditions used during learning. This would turn $\rho_{\tau}$ from a simulation parameter into an experimentally calibrated quantity.

The hybrid result can also be tested in hardware. The present architecture uses a fast-relaxing and a persistent branch contributing to the same logical weight, but a four-device differential synapse is only one possible implementation. Heterogeneous device pairs, multi-state devices combining volatile and non-volatile components, or mixed analog synapses with a local fast branch and a less frequently updated persistent branch could provide the same functional separation. Such implementations could reduce the twofold device-count penalty observed here while retaining multiple memory timescales.

The most direct hardware experiment would preserve the sequential protocol used in the simulations. A memristive SNN would first be programmed and trained on $S_1$ and then exposed chronologically to $S_2,\ldots,S_{10}$. Conductance trajectories, programming pulses, current-session accuracy, forgetting, and backward transfer would be measured after each adaptation stage. Repeating the sequence under controlled electrode displacement would allow the environmental timescale to vary independently and would test whether the high-performance region near comparable synaptic and input dynamics remains visible in physical hardware.


\section{Conclusion}
\label{sec:conclusion}

This paper considered continual myoelectric decoding in terms of two interacting temporal processes: the physical memory of a memristive synapse and changes in the physiological input. Six literature-constrained memristive device models were implemented in the same SNN and evaluated using an identical sequential NinaPro DB6 protocol. Despite the common network architecture and learning rule, their continual behavior differed markedly. Session-1 performance did not predict these differences. Devices capable of strong adaptation could also show substantial forgetting, while greater persistence improved retention without necessarily producing the highest continual-learning accuracy.

The controlled-timescale experiment separated synaptic relaxation from input drift. Across the $(\tau_{\mathrm{mem}},\tau_{\mathrm{drift}})$ space, the highest $BA_{\mathrm{CL}}$ occupied a broad region where the two temporal scales were comparable. The maximum, approximately $0.565$, occurred for $\rho_{\tau}=\tau_{\mathrm{mem}}/\tau_{\mathrm{drift}}$ slightly below unity. A model based on $\rho_{\tau}$ reached a leave-one-participant-out RMSE of $0.0154$, whereas models using $\tau_{\mathrm{mem}}$ or $\tau_{\mathrm{drift}}$ alone reached $0.0328$ and $0.0327$, respectively. The additive model containing both absolute timescales remained less predictive, with an RMSE of $0.0215$. Under the present protocol, synaptic retention was more informative when expressed relative to the rate of input change than when treated as an isolated device property.

The Ioannou--Sahu hybrid provided a practical test of this relation. Combining fast and persistent synaptic components increased $BA_{\mathrm{CL}}$ to $0.516$, compared with $0.439$ for Ioannou and $0.459$ for Sahu, while reducing the forgetting associated with the fast branch and retaining most of its adaptation capability. This gain required twice as many physical memristive elements per logical synapse and a larger cumulative programming burden. Experimental validation will require measurement of synaptic and sEMG drift timescales within the same physical system, together with testing under array-level nonidealities, device ageing, finite endurance, and longitudinal myoelectric variability.

%
%
%


\ack{
The development of the spiking neural network architecture 
was supported by the Ministry of Science and Higher 
Education of the Russian Federation under state assignment 
(project No.~FSMG-2025-0070). The investigation of memristive 
device properties and material-dependent timescales was 
supported by state assignment (project No.~FSWR-2025-0006) 
of the Ministry of Science and Higher Education of the 
Russian Federation. Computational modeling was supported 
by the Russian Science Foundation (project No.~25-71-30008).
}

\section*{Conflict of interest}

The author declares no competing financial or non-financial interests.

\section*{Ethics statement}

This study used previously published, de-identified data from the NinaPro DB6 dataset and involved no new recruitment or experimental procedures with human participants.

\data{
The data and source files supporting the findings of this paper are publicly available at \url{https://github.com/Kipelkin/sEMG}.
}

\suppdata{
Supplementary data are available with the online version of this article.
}


\bibliographystyle{unsrt}
\bibliography{references}

\end{document}